\documentclass[%
aip,
nofootinbib,
 amsmath,amssymb,
 reprint,
]{revtex4-1}
\usepackage{placeins}
\usepackage{amsmath}
\usepackage{float}
\usepackage{siunitx}
\usepackage[dvipsnames]{xcolor}

\usepackage{lipsum}
\usepackage{graphicx}
\usepackage{dcolumn}
\usepackage{bm}
\usepackage{graphicx}
\usepackage{dcolumn}
\usepackage{bm}
\usepackage[version=4]{mhchem}
\usepackage{amsmath}
\usepackage{empheq}
\usepackage{mdframed}
\usepackage{xcolor}
\usepackage{float}
\usepackage{subcaption}

\usepackage[utf8]{inputenc}
\usepackage[T1]{fontenc}
\usepackage{mathptmx}
\usepackage{etoolbox}

\begin{document}
\preprint{APS/123-QED}

\title{Post-Collision Thermal Excitation and Survival-Limited Cluster Growth in the Gas Phase}

\author{Tomoya Tamadate}
 \email{tamadate@se.kanazawa-u.ac.jp}
 \affiliation{%
 Faculty of Frontier Engineering,Institute of Science and Engineering, Kanazawa University}

\author{Christopher J. Hogan Jr.}%
\email{hogan108@umn.edu}
\affiliation{%
 Department of Mechanical Engineering, University of Minnesota}
 
%
%
\date{\today}

\begin{abstract}
Gas-phase cluster growth by monomer addition is commonly modeled as an isothermal process. We develop a survival-limited framework in which association produces a thermally excited cluster that may dissociate before either cooling through bath-gas collisions or encountering the next monomer. A continuous-energy survival probability is first derived for an individual post-association thermal trajectory and is then marginalized over distributions of excitation energy, equilibrium energy, and monomer-arrival time using a trajectory functional. Molecular-dynamics simulations of water, silver, and gold clusters provide size-dependent caloric relationships and latent heats, while event-based Monte Carlo simulations independently test the survival formulation. Theory and Monte Carlo results agree closely. The ensemble-averaged survival probability exhibits strong and non-monotonic size dependence, with the largest thermal penalties generally occurring for the smallest clusters. Intermediate-size local maxima arise only when complete cluster-energy distributions are retained and result from competition between curvature-enhanced dissociation and the narrowing of the low-energy tail with increasing size. Surviving clusters are consequently drawn preferentially from the colder portion of the pre-collision energy distribution, and mean thermal trajectories can substantially underestimate population survival. To connect single-event survival to cumulative growth, we introduce a thermal forward-rate correction relative to an isothermal reference and incorporate it into a reversible birth--death model. Although the correction at each size may be moderate, its multiplicative accumulation can increase mean first-passage times by many orders of magnitude. The framework provides a general route for identifying post-collision stabilization as a control on gas-phase cluster growth.
\end{abstract}
\maketitle


\section{Introduction}
Cluster growth in the gas phase commonly proceeds by monomer addition in scenarios wherein the cluster number density is small relative to the monomer density itself, i.e. in highly supersaturated systems \cite{Wyslouzil_2016}. While for supermicrometer droplets simultaneous heat and mass transfer are commonly considered in modeling growth \cite{Nadykto_2003}, a near-ubiquitous assumption in aerosol growth modeling at the cluster (nanometer) scale is that the growth process itself is isothermal \cite{Mcgrath_2012,FISK1998}, i.e. that all clusters are thermally equilibrated with the surrounding bath gas instantly, or that cluster energy fluctuations have a negligible influence on the growth process. For example, the nucleation rate, defined as the rate at which clusters form, is typically derived by balancing growth via monomer-cluster collisions (the forward association reaction) and monomer evaporation (the backward dissociation reaction) under assumed isothermal conditions \cite{Girshick_1990,Wedekind_2007,KATZ1977}. In the case of relatively non-volatile condensing monomers, isothermal growth leads to the assumption that growth occurs at the collision-controlled limit \cite{Rao_1989}. 

However, the isothermal assumption is not rigorously valid; as discussed in detail in the seminal work of Feder et al. \cite{Feder1966} monomer addition leads to latent heat release to a cluster, while dissociation correspondingly leads to cooling. Meanwhile the bath gas serves to thermally equilibrate the cluster on a finite timescale. These changes in cluster internal energy have a direct influence on the propensity for monomer dissociation. Feder et al. \cite{Feder1966} demonstrated that a detailed balance describing a cluster population should not only consider evolution of the number density of clusters composed of $g$ monomers via forward and backward reactions, but also via consideration of the internal energies $E$ of clusters, which are affected by latent to sensible transfer during association \cite{Freund_1977} and dissociation as well as collisional heat transfer (conduction) with the bath gas. Their analysis of this two-dimensional balance, which focused on nucleation rates, suggested the population of clusters below the critical size may be systematically cooler than the bath gas temperature because they are formed via dissociation reactions, stabilizing them against further dissociation, and which may ultimately influence formation and growth rates \cite{Toxvaerd_2015,Valtteri_2022}. Non-isothermal effects have been further examined and incorporated into models by others, though not always arriving at the same conclusion. Wyslouzil \& Seinfeld \cite{Wyslouzil_1992} derived nucleation rates considering that clusters had a size-dependent specific energy $g(E)$ where $E$ was distinct from the mean bath-gas equilibrated cluster energy. Notably, in multiple studies Barrett \& colleagues \cite{Barrett_1993,Barrett_1994,Barrett_2008, Barrett_2011, Ford_1989} have examined nucleation rates derived from two-dimensional balance solutions, which suggest that both sub-critical and super-critical clusters can be hotter than the surrounding bath gas, and that nucleation rates predicted with clusters not in equilibrium with their surroundings are an order of magnitude or more below isothermal predictions.\cite{Schweizer_2014} 

\begin{figure*}[t]
    \centering
    \includegraphics[width=\textwidth]{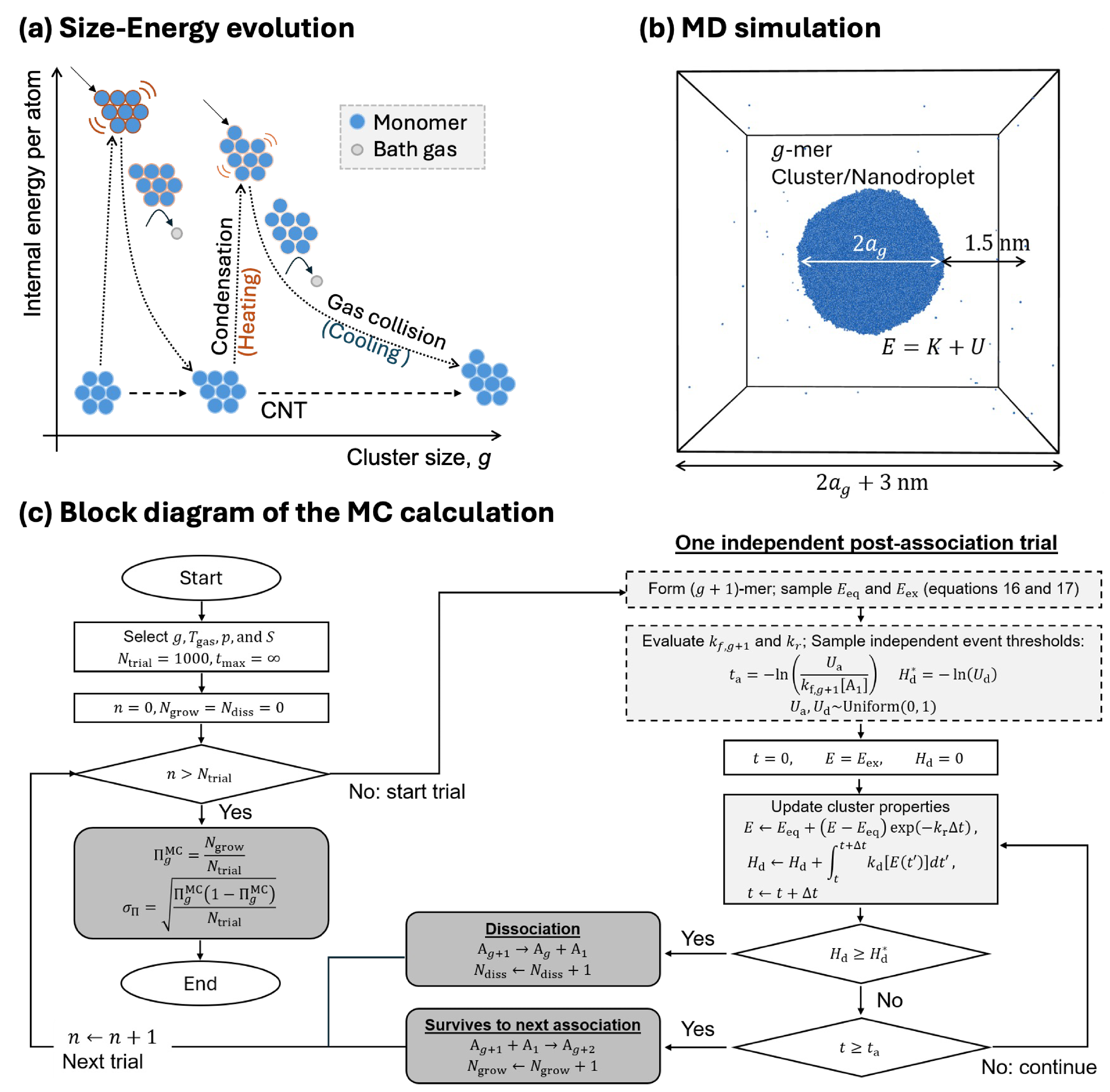}
    \caption{\label{fig:overview}Overview of the theoretical and computational framework proposed. (a) Cluster evolution in size--internal-energy space. Monomer association increases both cluster size and internal energy, bath-gas collisions cool the cluster, and dissociation competes with stabilization; the horizontal dashed path denotes the isothermal "CNT" description.
    (b) Molecular-dynamics simulation geometry used to determine the
    size-dependent cluster caloric relationship, $E=K+U$. (c) Event-based
    Monte Carlo calculation of whether a newly formed $(g+1)$-mer dissociates or survives until the next monomer association, and estimation of the ensemble-averaged survival probability $\Pi_g^{\rm MC}$. }
\end{figure*}

There remains continued interest in better understanding how the interplay between size and energy evolution in association reactions, dissociation reactions, and bath gas interactions not only affect nucleation rates, but also more generally how cluster populations evolve both in size and energy. Beyond examination of nucleation rates, Zachariah \& Colleagues \cite{ZACHARIAH1999, Zachariah_Carrier_Blaisten-Barojas_1994, Lehtinen_2001} as well as Yang et al. \cite{Yang_2019} showed that latent heat release during cluster-cluster collisional growth (coagulation) is even more significant than in cluster-monomer growth, and in atomistic simulations, Blaisten-Barojas \& Zachariah \cite{Blaisten-Barojas_1992} observed that the heat release during coalescence of two clusters can drive monomer dissociation. Cluster population evolution simultaneously considering size and energy changes due to collisional growth, dissociation, and bath gas interaction have been studied via constant number Monte Carlo techniques, which show rather distinct size distribution evolution behavior from collision-controlled growth models in the presence of a finite vapor concentration \cite{CHEN2024}. Ojha et al. \cite{Ojha_2024} developed a combined model to examine combined latent heat release during cluster-cluster coalescence and subsequent cooling by bath gas, demonstrating that growing clusters are transiently hotter than their surroundings, while Lalanne et al. \cite{LALANNE2022} experimentally inferred that growing nanoparticles are hotter than their surroundings in a flame synthesis system. However, combining prior studies focused on nucleation rates, and studies focused on heating during cluster-cluster growth (a separate process), there appears to be more limited prior effort \cite{Freund_1977,bauer_1977} to specifically derive relationships predicting how cluster-monomer growth rates are affected by latent heat transfer and interaction with the bath gas, i.e. to derive growth rate estimates which do not require solution to multidimensional detailed balances, and which account for thermal excitation of clusters leading to increased propensity for monomer dissociation. Such estimates would be of value in estimating how thermal excitation influences cluster growth. Towards improved understanding of non-isothermal influences on cluster growth via monomer addition, here we develop a relationship for the effective collisional growth rate of clusters considering that upon monomer association ($\ce{A_g + A_1 -> A_{g+1}}$ for cluster species $A$)\cite{Buckle_1969}, the newly formed cluster is in a thermally excited state, and is cooled by the surrounding bath gas at a finite rate. While in this excited state, the cluster dissociation rate is elevated relative to the equilibrium rate. To truly grow, the newly formed cluster therefore must "survive", i.e. not dissociate ($\ce A_{g+1}\rightarrow \ce A_g+A_1$) prior to the arrival of the next monomer ($\ce A_{g+1}\rightarrow \ce A_{g+2}$). Considering this reaction system (depicted in Figure \ref{fig:overview}(a) over $(g,E)$ space), our approach to arrive at a net forward growth rate $\frac{d[\ce A_{g+1}]}{dt}$ for the net reaction $\ce{A_g + A_1 -> A_{g+1}}$ in an aerosol is based on survival/hazard analysis. We consider separate reaction pathways for thermal relaxation by bath gas interaction and dissociation, until ultimately a subsequent monomer arrives. This results in a survival probability for a newly-formed cluster, and the product of the survival probability and collision rate represents the maximum rate at which clusters can grow. In the following sections, we derive the survival probability expression in detail, demonstrating a distinct exponential pressure dependence that differs fundamentally from prior discrete-collision models. Our derivation is first applied to a single "thermal trajectory", defining the relaxation of a cluster from a specific post-association excited state to a specific potential final energy state.  To determine an ensemble-averaged survival and net reaction rate for association we then apply a trajectory functional to integrate over an invariant distribution of continuous thermal histories. We show that for three test species, water, silver, and gold clusters (all of which have been examined in aerosol systems \cite{Wagner_1981,Brus_2009,Campagna_2020,MAISSER_2021, Feusi_2024}), thermal excitation leads to an upper limit estimate for the growth rate which is well below the traditional collision-controlled growth rate. Calculations additionally reveal a strong, non-monotonic size-dependency for the survival-limited growth rate for clusters in the nanometer scale size regime and suggest that the clusters which are more likely to survive (complete) the growth process are those sampled from the lower tail of the energy distribution.

\section{Theory \& Methods}
\subsection{Cluster Growth Survival Probability}
We consider the association reaction:

\begin{equation}
\ce{A_g + A_1 ->[\textit{k}_f] A^{\textit{i}_{max}}_{\textit{g}+1}}
\end{equation}

\noindent occurring in a bath gas at prescribed gas number density and temperature, where $g$ denotes the number of monomers in a cluster of species "$\ce A$" and the superscript $i$ denotes total energy level above the "final" energy level $i=0$. $i=i_{\rm max}$ corresponds to the maximally excited energy level, formed immediately following monomer incorporation, via conversion of latent heat to sensible heat. We first examine a single thermal "trajectory" ($\Gamma_T$), which is a single maximally excited energy level and a single final energy level. $k_f$ corresponds to the forward, collision-controlled reaction rate coefficient for a monomer and the cluster \cite{Yang_2018}. The newly formed, thermally excited cluster can then both thermally relax, and dissociate a monomer. We first consider energetic relaxation in discrete steps: 

\begin{subequations}
\begin{align} 
  \ce{A^{\textit{i}_{max}}_{\textit{g}+1} &->[\textit{k}^{\textit{i}_{max}}_{r}]& A^{\textit{i}_{max}-1}_{\textit{g}+1}} \\
  \ce{A^{\textit{i}}_{\textit{g}+1} &->[\textit{k}^{\textit{i}}_{r}] &A^{\textit{i}-1}_{\textit{g}+1}} \\
  \ce{A^{\textit{i}_{max}}_{\textit{g}+1} &->[\textit{k}^{\textit{i}_{max}}_{d}]& A_g + A_1} \\
  \ce{A^{\textit{i}}_{\textit{g}+1} &->[\textit{k}^{\textit{i}}_{d}]& A_g + A_1}.
\end{align}
\end{subequations}

\noindent Here, $k_r^{i}$ denotes the thermal relaxation rate associated with heat transfer between a cluster at excited thermal energy level $i$ and the bath gas, while $k_d^{i}$ denotes the corresponding thermal-energy-dependent dissociation rate. For simplicity, throughout this work we assume monotonic energy decay due to interaction with the bath gas. More rigorously, the cluster energy follows a stochastic relaxation trajectory, which in the diffusion limit may be represented by an Ornstein–Uhlenbeck-type process with fluctuations about the mean thermal relaxation path. Combining  elementary steps, the balance equations for all species are:

\begin{subequations}
\begin{align} 
&  \frac{d[\ce{A^{\textit{i}_{\rm max}}_{\textit{g}+1}}]}{dt}\Bigg|_{\Gamma_T} = k_f [\ce{A_g}][\ce{A_1}] - k^{i_{\rm max}}_r [\ce{A^{i_{\rm max}}_{g+1}}]-k_d^{i_{\rm max}}[\ce{A^{i_{\rm max}}_{g+1}}]=0\\
&  \frac{d[\ce{A^{\textit{i}}_{\textit{g}+1}}]}{dt}\Bigg|_{\Gamma_T} = k^{i+1}_{r} [\ce{A^{i+1}_{g+1}}] - k^{i}_r [\ce{A^{i}_{g+1}}]-k_d^{i}[\ce{A^{i}_{g+1}}]=0\\
&  \frac{d[\ce{A^0_{g+1}}]}{dt}\Bigg|_{\Gamma_T}=k^1_r[\ce{A^1_{g+1}}].
\end{align}
\end{subequations}

\noindent where $t$ denotes time, $[\ce{A^{\textit{i}}_{\textit{g}}}]$ denotes the number density of clusters with energy level $i$ composed of $g$ monomers, and the subscript $\Gamma_T$ denotes the relationship applies for a single thermal trajectory. Our goal is to develop an expression for the effective association rate $\frac{d[\ce{A^0_{g+1}}]}{dt}\Bigg|_{\Gamma_T}$ for a single thermal trajectory, considering that the newly formed cluster may dissociation before it thermally relaxes to its final state. We can invoke a quasi steady-state approximation (QSSA) for all excited species, considered to be short-lived reactive intermediates. This yields:

\begin{align}
\bigl[\mathrm{A}_{g+1}^{i_{\rm max}}\bigr]\Big|_{\Gamma_T}
 &= \frac{k_f \,\bigl[\mathrm{A}_g\bigr]\bigl[\mathrm{A}_1\bigr]}
     {k_{r}^{\,i_{\rm max}} + k_{d}^{\,i_{\rm max}}}\\
\bigl[\mathrm{A}_{g+1}^{i}\bigr]\Big|_{\Gamma_T}
 &= \frac{k_{r}^{\,i+1}\,\bigl[\mathrm{A}_{g+1}^{i+1}\bigr]}
     {k_{r}^{\,i} + k_{d}^{\,i}},
\end{align}

\noindent Combining equations then yields:
\begin{widetext}
\begin{align} \label{eq:product_sum}
  \frac{d[\ce{A^{0}_{\textit{g}+1}}]}{dt}\Bigg|_{\Gamma_T} = k^1_r\frac{k^2_r}{k^1_r+k^1_d}\frac{k^3_r}{k^2_r+k^2_d}...\frac{k^{i_{\rm max}}_r}{k^{i_{\rm max}-1}_r + k^{i_{\rm max}-1}_d} \frac{k_f[\ce{A_g}][\ce{A_1}]}{k^{i_{\rm max}}_r + k^{i_{\rm max}}_d} 
  = \left( \prod^{i_{\rm max}}_{i=1}\frac{k^{i}_r}{k^{i}_r + k^{i}_d} \right) k_f[\ce{A_g}][\ce{A_1}]
\end{align}
\end{widetext}

\noindent Equation (\ref{eq:product_sum}) is similar in structure and in derivation to that proposed by Bauer \& Frurip \cite{bauer_1977}, though Bauer \& Frurip consider a simpler scheme where one single collision with the bath gas is sufficient for rethermalization. This simplification yields a Lindemann–Hinshelwood-like \cite{laidler1965chemical} dependency on pressure, which we will show subsequently is distinct from predictions of the model developed here. The product chain can then be adjusted to a sum with an exponential:
\begin{equation}
  \frac{d[\ce{A^{0}_{\textit{g}+1}}]}{dt}\Bigg|_{\Gamma_T} 
  = k_f[\ce{A_g}][\ce{A_1}] \exp \left[
    \sum_{i=1}^{i_{\rm max}}\ln \left( \frac{k^{i}_r}{k^{i}_r + k^{i}_d}\right) \right]
  \label{eqn:exp_sum}
\end{equation}

\noindent We then note that steps in $i$, which is the coordinate along the thermal relaxation path, can be made infinitesimally small, allowing us to recast equation (\ref{eqn:exp_sum}) as:
\begin{equation}
  \frac{d[\ce{A^{0}_{\textit{g}+1}}]}{dt}\Bigg|_{\Gamma_T} 
  = k_f[\ce{A_g}][\ce{A_1}] \exp \left[
    \int_{i=0}^{i_{\rm max}}\ln \left( \frac{k^{i}_r}{k^{i}_r + k^{i}_d}\right) di \right] 
\end{equation}

\noindent We can then shift from integrating across energy levels to integrating across energy, via the relationships $\frac{dE}{dt'}=-k_r(E)(E-E_{\rm eq})$ and $\frac{di}{dt'}=-k_r^i$, where $k_r^i=(i_{\rm max}+i_\Delta)k_r(E)$, $i=i_{\rm max}$ corresponds to $E=E_{\rm ex}$, and $i_\Delta$ is the energy level difference between the final trajectory energy level $E_0$ and the gas equilibrium energy level $E_{\rm eq}$, hence $di=\frac{i_{\rm max}+i_\Delta}{(E-E_{\rm eq})}\,dE$. $E_{\rm eq}$ is distinct from the final energy level considered $E_0$ because the cluster requires an infinite amount of time to reach $E_{\rm eq}$ via bath gas interactions. Further noting that $i_{\rm max}$ is arbitrarily large ($i_{\rm max} \rightarrow \infty$) while $k_r(E)$ remains non-zero, using $\ln(1+x)\approx x$ for $x \ll 1$ yields:
\begin{align}
  \frac{d[\ce{A^{0}_{\textit{g}+1}}]}{dt}\Bigg|_{\Gamma_T} 
    = k_f[\ce{A_g}][\ce{A_1}] \exp \left(
    -\int_{E_{\rm 0}}^{E_{\rm ex}} 
    \left(\frac{1}{E-E_{\rm eq}}\right)
    \frac{k_d(E)}{k_r(E)}\,dE \right)
\end{align}

\noindent The relaxation rate and the dissociation rate are first-order reactions, characterized by first-order reaction times ($\tau_{\rm r}(E)$ and $\tau_{\rm d}(E)$), respectively, leading to:
\begin{equation} \label{eq:singleIntegralCoeff}
  \frac{d[\ce{A^{0}_{\textit{g}+1}}]}{dt}\Bigg|_{\Gamma_T} 
    = k_f[\ce{A_g}][\ce{A_1}] \exp \left(-\int_{E_{\rm 0}}^{E_{\rm ex}} 
    \left(\frac{1}{E-E_{\rm eq}}\right) \frac{\tau_{\rm r}(E)}{\tau_{\rm d}(E)}\,dE \right)
\end{equation}

\noindent Before proceeding further we remark that equation (\ref{eq:singleIntegralCoeff}) suggests unique pressure dependency for the growth rate which is additionally distinct from the predictions of Bauer \& Frurip \cite{bauer_1977}. Shown subsequently $\tau_r$ is inversely proportional to the bath gas pressure. In the high pressure limit, $\frac{d[\ce{A^{0}_{\textit{g}+1}}]}{dt}\Big|_{\Gamma_T}\propto k_f[\ce{A_g}][\ce{A_1}](1-\frac{C^*_{1}}{P})$, where $C_{*1}$ is a constant and $P$ is pressure; this is similar to Lindemann-Hinshelwood kinetics and suggests a diminished influence of pressure as the reaction approaches the collision-controlled limit (as $P\rightarrow\infty$). However, outside the high pressure limit the approach here yields $\frac{d[\ce{A^{0}_{\textit{g}+1}}]}{dt}\Big|_{\Gamma_T}\propto k_f[\ce{A_g}][\ce{A_1}]\exp(-\frac{C^*_{2}}{P})$. This distinct behavior is a direct consequence of assuming continuous energy levels for a cluster and a continuous relaxation path. 

Equation (\ref{eq:singleIntegralCoeff}) also has the form $\frac{d[\ce{A^{0}_{\textit{g}+1}}]}{dt}\Big|_{\Gamma_T}=k_f[\ce{A_g}][\ce{A_1}]S(E_{\rm ex},E_{\rm eq},E_{\rm 0})$ where $S(E_{\rm ex},E_{\rm eq},E_{\rm 0})$ is the survival probability for a cluster, defined as the probability that the cluster does not dissociate prior to thermally relaxing to the final energy $E_{\rm 0}$. The survival-transform structure of our rate equation is similar to functional forms which arise for the stabilization of molecules in unimolecular reaction theory \cite{slater1960theory, Malpathak_2019}, though here it is constructed mechanistically from continuous thermal relaxation of a newly formed cluster. With the previously noted definition of $\frac{dE}{dt'}$, the survival probability can be equivalently written in its more typical form as $S(E_{\rm ex},E_{\rm eq},E_{\rm 0})=\exp \left(-\int_0^{t_{\rm f}} \tau^{-1}_{\rm d}(E(t'))dt' \right)$ for a first order reaction, which again identifies that it depends upon a clearly defined final time $t_{\rm f}$, an equation describing the dissociation rate, and the energy time history during the thermal relaxation process. Though not examined here, this form would additionally enable implementation with non-monotonic energy evolution for a given thermal trajectory. 

To define a final time for the survival probability, we first remark that as $t_{\rm f}\rightarrow\infty$, $S(E_{\rm ex},E_{\rm eq},E_{\rm 0})\rightarrow0$ unless $\tau_{\rm d}\rightarrow\infty$. We then note that growth itself is a continuous process, and that while a thermally excited cluster relaxes, it will eventually encounter another monomer, facilitating further growth. Therefore, an appropriate final time measure is $t_{\rm f}=t_{\rm a}$, where $t_{\rm a}$ is the arrival time of the next monomer (i.e. the time at which the reaction $A_{g+1}+A\rightarrow A_{g+2}$ occurs). The survival probability for a single thermal trajectory therefore is a function of the excitation energy, the equilibrium energy, and the arrival time of a subsequent monomer, $S(E_{\rm ex},E_{\rm eq},t_{\rm a})$. 

\subsection{Marginalization over Trajectory Distributions}
We return to initial assumption of a single thermal trajectory, defined by an initial excited energy, an arrival time for a subsequent monomer, an equilibration energy. The latter two define the final energy for the trajectory, and the equilibration energy also varies because cluster energies in thermal equilibrium themselves are broadly distributed. To better capture the dynamics of cluster populations, the influence of all possible thermal trajectories on the survival must now be considered. Rigorously, cluster-monomer collisions occur over a distribution of velocities and impact parameters; however, we assume that the influence of these parameters on the survival is small, as the larger influence is the latent heat release associated with monomer binding upon collision. Instead, thermal trajectories arise from "packets" of trajectories $G(\Gamma_T)d\Gamma_T$, where $G(\Gamma_T)$ represents the probability density function of trajectories contributing to growth.  For closure, we propose $G(\Gamma_T)d\Gamma_T=f_{\rm ex}(E_{\rm ex})f(t_{\rm a})f(E_{\rm eq})dE_{\rm ex}dt_{\rm a}dE_{\rm eq}$, where $f$ denotes the probability density function for each parameter. The ensemble-averaged survival (hereafter $\Pi_g$) over the distribution of all possible trajectories can then be computed via a thermal trajectory functional:
\begin{equation} \label{eqn:trajectory_functional}
  \Pi_g=\mathcal{S}[G]=\int_{\Omega_\Gamma}S(\Gamma_T)G(\Gamma_T)d\Gamma_T
\end{equation}

\noindent With $\Omega_\Gamma$ the range of all possible trajectory parameters, $\mathcal{S}[G]$ is a functional which converts the survival probability for a single trajectory into $\Pi_g$ considering all trajectories. For $\tau_r$ now modeled as independent of cluster energy, $E_0=E_{eq}+(E_{\rm ex}-E_{\rm eq})\exp(-\frac{t_{\rm a}}{\tau_r})$, and the trajectory functional leads to the survival-corrected association rate considering all trajectories, $\frac{d[\ce{A_{g+1}}]}{dt}$:

\begin{subequations}
    \begin{align}
&\frac{d[\ce{A_{g+1}}]}{dt}
  = k_{f,g}[\ce{A_g}][\ce{A_1}]\Pi_g\\
&\Pi_g=
  \int_0^\infty
  \int_0^\infty
  \int_0^\infty S(E_{\rm ex,},E_{\rm eq},t_a)
    f_{\rm ex}(E_{\rm ex})f(t_{\rm a})f(E_{\rm eq})
    \, dE_{\rm ex}\,dt_a\,dE_{\rm eq} \\
   &S(E_{\rm ex,},E_{\rm eq},t_a) = 
\exp\!\left[-\displaystyle\int_{E_{eq}+(E_{\rm ex}-E_{\rm eq})\exp(-\frac{t_a}{\tau_r})}^{E_{\rm ex}}
\left(\frac{1}{E-E_{\rm eq}}\right)
\dfrac{\tau_r(E)}{\tau_d(E)}\, dE
\right]
\label{eq:central_eqn}     
\end{align}
\end{subequations}

What remains is to then define the distributions of thermal trajectories, as well as estimates of dissociation and thermal relaxation time constants. These are addressed subsequently. We remark that in defining the bounds of the trajectory functional, we have assumed $E_{\rm ex}$ and $E_{\rm eq}$ are completely independent of one another; $E_{\rm eq}$ is an equilibrated state driven by the bath gas, while $E_{\rm ex}$ results from the energy distribution originally in $\ce {A_g}$ clusters. There is a possibility that $E_{\rm ex}<E_{\rm eq}$. However, equation (\ref{eq:central_eqn}) need not be modified to address this situation; dissociation as the cluster equilibrates still must be considered to examine cluster survival even in instances where thermal relaxation involves heating to a higher thermal energy.

\subsection {Distributions \& Rates Defining Thermal Trajectories}
Three distributions and two characteristic times are required in $\Pi_g$ estimation. Of these, arguably the most unambiguous is the arrival time distribution for the subsequent collision with a vapor monomer $f(t_{\rm a})$, which is exponentially distributed\cite{Tamadate_2025}:
\begin{equation} \label{eqn:fta}
  f(t_{\rm a})=k_{f,g+1}[\ce{A_1}]\exp\left(-k_{f,g+1}[\ce{A_1}]t_{\rm a}\right)
\end{equation}

\noindent where the association rate coefficient is evaluated at the collision-controlled (free molecular) limit:
\begin{equation} \label{eqn:kf}
  k_{f,g+1} = \pi a_{g+1}^2\left(\frac{8k_{\rm B}T_{\rm gas}}{\pi m_{\rm A}}\right)^{1/2}
\end{equation}

\noindent Here, $k_{\rm B}$ is the Boltzmann constant, $m_{\rm A}$ is the monomer mass, and $a_{g+1}=\left[3(g+1)v_{\rm m}/4\pi\right]^{1/3}$ is the cluster radius, with $v_{\rm m}$ the condensed-phase molecular volume. We neglect the influence of the monomer finite size in this calculation as well as the enhancement in calculation rate due to potential interactions\cite{Yang_2018}; both of these shift the arrival time distribution to shorter times but outside of the smallest clusters, these influences are small. The monomer number density is prescribed via the saturation ratio $S_{\rm m}$, i.e. $[\ce{A_1}]=S_{\rm m}\,p_{\rm sat}(T_{\rm gas})/k_{\rm B}T_{\rm gas}$, where $p_{\rm sat}$ is the (flat-interface) saturation vapor pressure. The mean arrival time is correspondingly $\langle t_{\rm a} \rangle = \left(k_{f,g+1}[\ce{A_1}]\right)^{-1}$.

More challenging is the excited energy distribution.  This distribution should arise from a shift to the energy distribution for the $\ce {A_g}$.  However, prediction of this distribution would require full solution to the partial differential equation both cluster size and energy, drastically increasing complexity, and arguably obviating the need for survival-limited rate estimation. To facilitate calculations, we assume that $\ce {A_g}$ clusters are equilibrated with the surrounding bath gas prior to collision. This assumption enables direct estimates of a baseline survival-limited growth rate, as the equilibrium and excited energy distributions then follow from the statistical mechanics of the cluster internal energy, linked to a caloric relationship between the cluster kinetic and potential energies. For a cluster composed of $g+1$ monomers, each monomer contributes $\nu$ kinetic degrees of freedom ($\nu=6$ for a rigid nonlinear water molecule and $\nu=3$ for a monatomic silver or gold vapor), and the total kinetic energy $K$ of a canonical-ensemble cluster at bath gas temperature $T_{\rm gas}$ is gamma distributed \cite{Bussi_2007}:
\begin{equation} \label{eqn:fK}
  f_{K}(K)=\frac{\left(K/k_{\rm B}T_{\rm gas}\right)^{\kappa-1}}{k_{\rm B}T_{\rm gas}\,\Gamma(\kappa)}\exp\left(-\frac{K}{k_{\rm B}T_{\rm gas}}\right)
\end{equation}

\noindent where $\kappa=(g+1)\nu/2$ is the gamma-distribution shape parameter. Molecular dynamics simulations of the test clusters examined here (described subsequently) reveal that away from phase transitions, the ensemble-averaged cluster potential energy $<U>$ is a linear function of the ensemble-averaged kinetic energy, $<U> = a<K>+b_{g+1}$, where $a\equiv a_{g+1}$ and $b_{g+1}$ are size- and material-dependent coefficients. Neglecting fluctuations about this linear relationship (which diminish in relative magnitude with increasing $g$), the total cluster energy is $E=K+U=(1+a)K+b_{g+1}$, and the change of variables $K=(E-b_{g+1})/(1+a)$ applied to equation (\ref{eqn:fK}) yields the equilibrium total energy distribution:
\begin{align} \label{eqn:feq}
  f_{\rm eq}(E_{\rm eq})={}&\frac{1}{(1+a)k_{\rm B}T_{\rm gas}\,\Gamma(\kappa)}
  \left[\frac{E_{\rm eq}-b_{g+1}}{(1+a)k_{\rm B}T_{\rm gas}}\right]^{\kappa-1} \notag \\
  &\times\exp\left[-\frac{E_{\rm eq}-b_{g+1}}{(1+a)k_{\rm B}T_{\rm gas}}\right]
\end{align}

\noindent defined for $E_{\rm eq}\ge b_{g+1}$. The linear caloric relationship also directly links the cluster energy to an effective cluster temperature, $E=\frac{(g+1)\nu}{2} k_{\rm B}(1+a)T_{\rm l}+b_{g+1}$, hence $T_{\rm l}(E)=2(E-b_{g+1})/\left[(g+1)\nu k_{\rm B}(1+a)\right]$, and implies a specific heat $c_{\rm p}=\nu k_{\rm B}(1+a)/(2m_{\rm A})$, which serves as a consistency check on MD-determined $a$ values against bulk property data. 

For the excited state formed immediately upon monomer incorporation, the newly formed $(g+1)$-mer carries the energy of its precursors; this energy is augmented above $(g+1)$-mer ensemble average equilibrium level by the latent heat $L_g(T)$. Approximating the pre-collision thermal energy content by the equilibrium distribution of the parent cluster $f_{{\rm eq},g}$, the excited energy distribution is the equilibrium distribution displaced by the latent heat:
\begin{equation} \label{eqn:fex}
  f_{\rm ex}(E_{\rm ex})=f_{{\rm eq},g}(E_{\rm ex}-L_g(T_{\rm gas})
\end{equation}

\noindent The latent heat itself is the average energy that must be transferred to the gas from the product cluster for thermal equilibration \cite{Yang_2019}. On a molar basis, the energy balance used here for attachment of a monomer to a $g$-mer is
\begin{subequations} \label{eqn:latent}
\begin{align}
L_g(T)={}&b_g+b_1-b_{g+1}-\Xi_g(T),\\
\Xi_g(T)={}&-RT\left\{\frac{1}{2}\right. \notag\\
&+\frac{1}{2}\left[(\nu g-3)a_g+(\nu-3)a_1
-\left(\nu(g+1)-3\right)a_{g+1}\right] \notag\\
&\left.-\frac{5g}{2(g+1)^2}\right\}.
\end{align}
\end{subequations}
$L_g(T)$ is thus calculated from the MD-derived caloric parameters $a_g$, $a_{g+1}$, and $b_g$.The temperature-dependent correction $\Xi_g(T)$ is generally small compared with the bond-energy difference $b_g+b_1-b_{g+1}$ and arises from the change in cluster heat capacity together with the conversion of collision translational energy into internal and rotational energy. Equation (\ref{eqn:latent}) reduces exactly to equation (11a) of Yang et al.\ \cite{Yang_2019} for monatomic species ($\nu=3$), i.e. silver and gold examiner here. Its use for species such as water is a rigid-molecule extension; after removal of whole-cluster translation, a $g$-mer has $\nu g-3$ kinetic degrees of freedom. For rigid TIP3P water, $\nu=6$, intramolecular vibrations are constrained, and the isolated-monomer potential energy is temperature independent, so $a_1=0$. We also use $b_1=0$ as the monomer potential-energy reference. The extension of Yang et al \cite{Yang_2019} to polyatomic species and its assumptions are derived in the Supplementary Material.  Equations (\ref{eqn:feq}) and (\ref{eqn:fex}) also show that association shifts the mean cluster temperature upward by $\Delta T = 2L_g/\left[(g+1)\nu k_{\rm B}(1+a_{g+1})\right]$; because $\Delta T \propto (g+1)^{-1}$ while the width of the temperature distribution scales as $(g+1)^{-1/2}$, latent heating displaces small clusters far into the high-energy tail of the equilibrium distribution, but this shift is buffered by heat capacity for larger clusters.

The thermal relaxation time $\tau_{\rm r}$ arises from a lumped-capacitance description of free molecular heat conduction between the cluster and the bath gas \cite{YANG2022}. Balancing the kinetic energy flux carried by impinging and scattered bath gas molecules with thermal accommodation coefficient $\alpha$ leads to $\frac{dE}{dt}=-\left(E-E_{\rm eq}\right)/\tau_{\rm r}$ with:
\begin{equation} \label{eqn:taur}
  \tau_{\rm r}=\frac{(g+1)m_{\rm A}c_{\rm p}}{2\pi a_{g+1}^2 k_{\rm B}\alpha}
  \left(\frac{\pi m_{\rm gas}}{8k_{\rm B}T_{\rm gas}}\right)^{1/2}\frac{k_{\rm B}T_{\rm gas}}{p}
\end{equation}

\noindent where $m_{\rm gas}$ is the bath gas molecule mass and $p$ is the bath gas pressure. We again neglect finite size effects of the gas molecule radius and the influence of gas molecule-cluster potential effects on approaching gas molecule trajectories to a particle. $\tau_{\rm r}$ is independent of the instantaneous cluster energy, consistent with the exponential relaxation trajectory invoked in section II.B, and is inversely proportional to pressure; it is this proportionality, inserted into the survival integral, that produces the $\exp(-C_2^*/P)$ pressure dependence noted in section II.A. For simplicity, we model the bath gas as diatomic nitrogen at atmopsheric pressure ($P_{gas}=10^5$ Pa) for all conditions, and we take $\alpha=1$ in all calculations; smaller accommodation coefficients lengthen $\tau_{\rm r}$ proportionally.

Finally, the dissociation time is modeled by treating monomer evaporation as a Hertz-Knudsen flux evaluated at the instantaneous cluster temperature $T_{\rm l}(E)$, with the Kelvin effect accounting for surface curvature:
\begin{align} \label{eqn:taud}
  \tau_{\rm d}(E)={}&\left[\frac{m_{\rm A}k_{\rm B}T_{\rm l}}{8\pi}\right]^{1/2}
  \frac{1}{a_{g+1}^2\,p_{\rm sat}(T_{\rm l})} \notag \\
  &\times\exp\left(-\frac{2\sigma v_{\rm m}}{k_{\rm B}T_{\rm l}a_{g+1}}\right)
\end{align}

\noindent where $\sigma$ is the (bulk) surface tension. Because $p_{\rm sat}$ depends exponentially on temperature (Clausius-Clapeyron scaling), modest thermal excitation can shorten $\tau_{\rm d}$ by orders of magnitude, coupling latent heat release to dissociation propensity. Combining equations (\ref{eqn:taur}) and (\ref{eqn:taud}) gives the hazard ratio appearing in the survival integral:
\begin{align} \label{eqn:ratio}
  \frac{\tau_{\rm r}}{\tau_{\rm d}}={}&\frac{(g+1)m_{\rm A}c_{\rm p}}{2k_{\rm B}\alpha}
  \left(\frac{m_{\rm gas}T_{\rm gas}}{m_{\rm A}T_{\rm l}}\right)^{1/2} \notag \\
  &\times\frac{p_{\rm sat}(T_{\rm l})}{p}
  \exp\left(\frac{2\sigma v_{\rm m}}{k_{\rm B}T_{\rm l}a_{g+1}}\right)
\end{align}

\noindent Equation (\ref{eqn:ratio}) shows that survival is governed by the saturation pressure evaluated at the excited cluster temperature relative to the bath gas pressure; the newly formed cluster is vulnerable while $p_{\rm sat}(T_{\rm l})$ remains comparable to or larger than $\alpha p$, and stabilizes as cooling drives $p_{\rm sat}(T_{\rm l})$ downward. For the assumed excited state distribution, evaluation of $\Pi_g$ therefore requires only the bath gas conditions ($T_{\rm gas}$, $p$, $S$), bulk property correlations for $p_{\rm sat}(T)$ and $\sigma(T)$, and the caloric coefficients $a$ and $b_{g+1}$.

\subsection{Thermal Properties of Test Clusters}
The caloric coefficients $a$ and $b_{g+1}$ linking cluster kinetic, potential, and total energies were determined from molecular dynamics (MD) simulations of isolated water, silver, and gold clusters, performed with LAMMPS \cite{Thompson_2022}, as depicted in Figure \ref{fig:overview}(b). All simulations placed a single cluster at the center of a periodic cubic cell whose edges extended at least $1.5$ nm beyond the cluster surface (vacuum environment), applied a Nos\'{e}-Hoover thermostat \cite{Nose_1984,Hoover_1985} with a $100$ fs damping time to all atoms, and used a $2$ fs integration time step. To sample the caloric curve, the thermostat set point was increased in a staircase protocol, holding the cluster at each isotherm for a fixed sampling window before stepping the temperature upward. Cluster center-of-mass motion was removed by recentering the largest cluster in the domain every $200$ fs. The total kinetic energy $K$, potential energy $U$, and instantaneous temperature were recorded every $200$ fs.

Water clusters were modeled with the rigid TIP3P potential \cite{Jorgensen_1983}, with O-H bonds and H-O-H angles constrained via the SHAKE algorithm \cite{Ryckaert_1977} and electrostatics evaluated by the Wolf summation method \cite{Wolf_1999} ($\alpha=0.1$ \AA$^{-1}$, $r_{\rm c}=15$ \AA). Clusters were generated by random insertion of molecules (minimum intermolecular spacing $1.5$ \AA) followed by energy minimization; the realized cluster sizes were $g=107$, $408$, $2,188$, $6,006$, $17,511$, $59,101$, and $140,091$ molecules. Each cluster was ramped from $250$ K to $350$ K in ten $10$ K isotherms, with per-isotherm sampling windows of $10$ ns ($g=107$), $1$ ns ($g=408$-$6,006$), $0.5$ ns ($g=17,511$), and $0.2$ ns ($g \ge 59,101$). Silver and gold clusters were modeled with embedded-atom-method potentials \cite{Foiles_1986}, and were generated by carving approximately spherical clusters from fcc lattices (lattice constants $4.09$ \AA \ and $4.08$ \AA, respectively). The extended metal size sweeps span $g=13$-$28,867$ for silver and $g=13$-$28,897$ for gold. Each size was sampled at 21 temperatures from $200$ K to $1200$ K in $50$ K increments using a $2$ fs time step, with $200{,}000$ equilibration steps followed by $800{,}000$ sampling steps per temperature.

\begin{figure*}
    \centering
    \includegraphics[width=\textwidth]{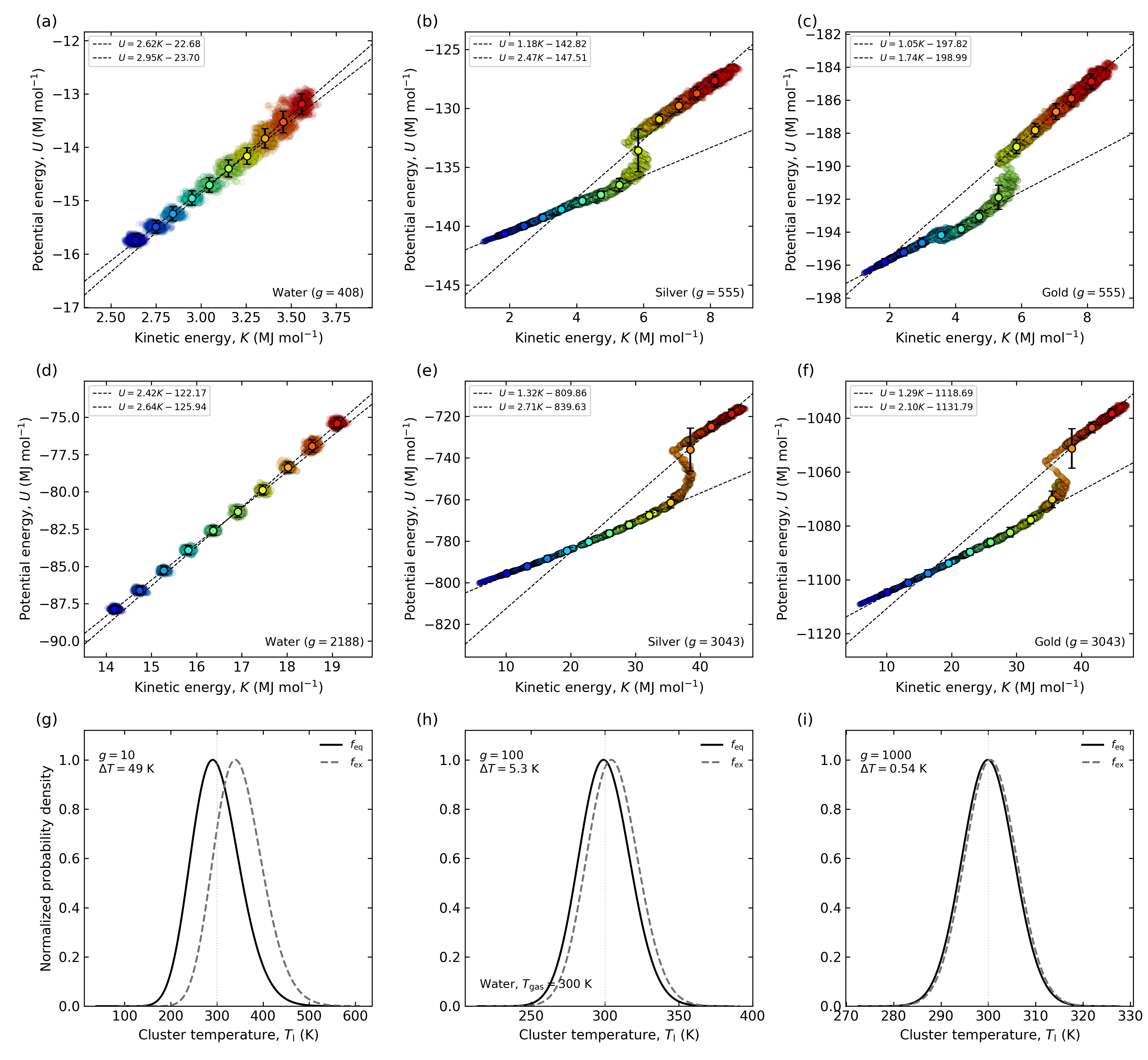}
    \caption{(a)-(f) MD caloric curves in the $U$-$K$ plane for two representative cluster sizes of water, silver, and gold. Small points are individual samples colored by isotherm; filled circles are isotherm block averages with $\pm 1 \sigma$ error bars on $U$; dashed lines are phase-specific linear fits $U=aK+b_{g+1}$. Melting appears as a discontinuous rise in $U$ at nearly constant $K$ for the metal clusters. (g)-(i) Equilibrium ($f_{\rm eq}$, solid) and post-association excited ($f_{\rm ex}$, dashed) energy distributions for water clusters at $T_{\rm gas}=300$ K, expressed on the cluster temperature axis, for $g=10$, $100$, and $1000$. The latent-heat displacement $\Delta T\propto (g+1)^{-1}$ decreases faster with increasing size than the distribution width $\propto (g+1)^{-1/2}$.}
    \label{fig:caloric}
\end{figure*}

For each isotherm, block averages $\langle K \rangle$ and $\langle U \rangle$ were computed after discarding the initial $2\%$ of the sampling window. Figure \ref{fig:caloric}(a)-(f) displays the resulting caloric curves in the $U$-$K$ plane for two representative cluster sizes of each material, along with the instantaneous (per-sample) values, which illustrate the magnitude of energy fluctuations about the caloric curve. In all cases the ensemble-averaged potential energy is well described by the linear relationship $\langle U\rangle=a\langle K\rangle+b_{g+1}$ within a single phase, supporting the change of variables leading to equations (\ref{eqn:feq}) and (\ref{eqn:fex}). For the metal clusters, melting is evident as a discontinuous rise in $U$ at nearly constant $K$; we therefore fit the solid and liquid branches separately, partitioning at the largest inter-isotherm jump in $U$. Water clusters remain liquid across the examined temperature range (no discernible phase transition, as expected given melting point depression at these sizes), and fits were restricted to the $260$-$310$ K isotherms, where evaporative loss during the sampling window is negligible. The continuous size-dependent parameterization obtained from the extended size sweeps is summarized in Table \ref{tab:caloric}.

\begin{table*}[t]
\caption{\label{tab:caloric} Production size-dependent caloric model $a_g=a_{\infty}+a_*g^{-1/3}$ and $b_g=b_0g$. The derived latent heat is shown at the lower fit boundary and in the large-size limit, and compared with the tabulated bulk value. Latent heats are evaluated at the listed reference temperature.}
\begin{ruledtabular}
\begin{tabular}{lcccccccc}
Material & $\nu$ & fitted $g$ & $a_{\infty}$ & $a_*$ & $b_0$ & $T_{\rm ref}$ & $L_{g_{\min}}\rightarrow L_{\infty}$ & $L_{\rm bulk}$ \\
 & & & & & (kJ mol$^{-1}$) & (K) & (kJ mol$^{-1}$) & (kJ mol$^{-1}$) \\
\hline
Water  & 6 & 107--140,091 & 2.2635 & 3.1653 & $-56$  & 200  & 43.28$\rightarrow$45.54 & 40.752 \\
Silver & 3 & 3043--28,867 & 1.0981 & 0.4360 & $-250$ & 1000 & 240.21$\rightarrow$240.46 & 255 \\
Gold   & 3 & 3043--28,897 & 1.0810 & 0.4604 & $-370$ & 1000 & 360.40$\rightarrow$360.67 & 321 \\
\end{tabular}
\end{ruledtabular}
\end{table*}

The per-monomer offset $b_{g}/g$ approaches a material-specific plateau, and we represent the fit intercepts by $b_g=b_0g$ and do not introduce a further size correction to $b_0$. The fit slopes retain a clear residual size dependence associated with the finite surface fraction. The specific heats implied by these slopes approach the bulk values with increasing size; rigid TIP3P modestly overestimates the bulk water heat capacity. The derived latent heats also differ from the tabulated bulk values because they retain the caloric model and potential-specific energy offset used in the simulations. For a compact three-dimensional cluster, the fraction of atoms or molecules at the surface scales to leading order as $g^{-1/3}$. We therefore fit the single-phase MD slopes to
\begin{equation} \label{eqn:afit}
a_g=a_{\infty}+a_* g^{-1/3}.
\end{equation}
All seven water cluster sizes ($107\leq g\leq140,091$) were included. For silver and gold, the slopes were evaluated over the common low-temperature window $200$-$500$ K. Small metal clusters exhibit structural and phase-dependent irregularities, so the production fit was restricted to the smooth asymptotic branch: seven sizes over $3,043\leq g\leq28,867$ for silver and $3,043\leq g\leq28,897$ for gold. The excluded data, independent structural diagnostics, and the sensitivity of the fitted form to alternative size corrections are documented in the Supplementary Material. The fit $(a_{\infty},a_*)$ values are $(2.2635,3.1653)$ for water, $(1.0981,0.4360)$ for silver, and $(1.0810,0.4604)$ for gold. Leave-one-out cross-validation against a constant model and alternative $g^{-1/2}$ and $g^{-2/3}$ corrections gave the lowest prediction error for equation (\ref{eqn:afit}) for all three materials. The theoretical calculations and Monte Carlo simulations use this same fitted $a_g$, the plateau representation $b_g=b_0g$, and the resulting size- and temperature-dependent latent heat $L_g(T)$ from equation (\ref{eqn:latent}); tabulated bulk values are retained only as reference comparisons.  
With equation (\ref{eqn:latent}) determined latent heats, exemplary $f_{\rm eq}$ and $f_{\rm ex}$ and distributions for water clusters are shown in figure \ref{fig:caloric}(g-i).  Distributions are expressed in terms of cluster temperature, calculated as: $T_{l}(E)=2(E-b_{g+1})/\left[(g+1)\nu k_{\rm B}(1+a)\right]$.  Because the latent-heat effect leads to shift in distributions $\Delta T\propto (g+1)^{-1}$ which decreases faster with increasing size than the distribution width $\propto (g+1)^{-1/2}$, latent heat effects are dampened out at larger cluster sizes.  Fitting caloric properties in this manner, i.e. extrapolating properties to smaller sizes, ignores any possible unique thermal properties for small clusters which may manifest (e.g. so-called "magic number" clusters \cite{Yamada_1992,Coolbaugh_1992}).  However, in an initial presentation and test of a survival-limited growth model, we find it is of greater importance to show that the model accurate recovers results from Monte Carlo simulations where both theory and simulation use identical cluster property models, as then the model framework can be applied later to examine cluster chemistry-specific behavior.  Further justification for small cluster exclusion in fitting is provided in the Supplementary Material. We also note that even with extrapolation, we perform calculations only beginning at $g=2$.  We find that thermal excitation is extremely pronounced for the reaction $A_1+A_1\rightarrow A_2^i$ such that survival is an extremely rare event in Monte Carlo simulations.  This does suggest that particular size-dependent and chemistry-specific treatment is needed at this scale and would certainly influence overall growth rates, as well as comparison between theory and experiment. 

\subsection{Event-Based Monte Carlo Validation}

To test the survival probability derivation and the trajectory-functional marginalization through a separate numerical route, we performed event-based Monte Carlo (MC) simulations which directly mimic the fate of individual post-association clusters. The MC simulations share the rate expressions (equations \ref{eqn:kf}, \ref{eqn:taur}, and \ref{eqn:taud}), the energy distributions (equations \ref{eqn:feq} and \ref{eqn:fex}), and the deterministic exponential cooling trajectory with the theory, but invoke none of the derivation steps of sections II.A and II.B; in particular, the QSSA is not applied,  and the survival integral is never evaluated. Agreement between MC and equation (\ref{eq:central_eqn}) therefore provides an internal test of the mathematical structure of the survival-probability formulation, rather than an independent validation of the shared physical rate models.

The MC method is depicted schematically in Figure \ref{fig:overview}(c). Each MC trial simulates a single newly formed $(g+1)$-mer as follows. First, the equilibrium and excited energies are sampled independently, $E_{\rm eq}=b_{g+1}+(1+a_{g+1})k_{\rm B}T_{\rm gas}X_1$ and $E_{\rm ex}=b_{g+1}+L_g(T_{\rm gas})+(1+a_{g+1})k_{\rm B}T_{\rm gas}X_2$, where $X_1$ and $X_2$ are independent gamma variates with shape parameter $\kappa$, consistent with equations (\ref{eqn:feq}) and (\ref{eqn:fex}). Second, the arrival time of the next monomer is sampled from equation (\ref{eqn:fta}) via an inverse transform, $t_{\rm a}=-\ln U_{\rm a}/\left(k_{f,g+1}[\ce{A_1}]\right)$, with $U_{\rm a}$ a uniform random variate on $(0,1)$. Third, dissociation is treated as a non-homogeneous Poisson process with time-dependent rate $\tau_{\rm d}^{-1}(E(t))$ evaluated along the cooling trajectory $E(t)=E_{\rm eq}+(E_{\rm ex}-E_{\rm eq})\exp(-t/\tau_{\rm r})$: a hazard threshold $H^{*}=-\ln U_{\rm d}$ is drawn, and the cumulative hazard $H(t)=\int_0^{t}\tau_{\rm d}^{-1}(E(t'))\,dt'$ is accumulated by trapezoidal integration with adaptive time steps $\Delta t = 0.05/\max(\tau_{\rm r}^{-1},\,k_{f,g+1}[\ce{A_1}],\,\tau_{\rm d}^{-1})$, interpolating the event time within the step in which $H$ crosses $H^{*}$ \cite{Gillespie_1977}. Once $t>30\tau_{\rm r}$, the cluster energy is within a factor of $10^{-13}$ of $E_{\rm eq}$ and the dissociation rate is treated as constant, allowing the remaining competition to be resolved analytically; no artificial time cutoff is imposed. The trial terminates in "growth" if the sampled arrival time is reached first ($t_{\rm a}<t_{\rm d}$), and in "dissociation" otherwise, exactly mirroring the survival competition underlying equation (\ref{eq:central_eqn}).

The MC growth probability is estimated as $\Pi_g^{\rm MC}=N_{\rm grow}/N_{\rm trial}$ from $N_{\rm trial}=1000$ independent trials per condition, with binomial standard error $\left[\Pi_g^{\rm MC}(1-\Pi_g^{\rm MC})/N_{\rm trial}\right]^{1/2}$. Simulations were performed at $50$ logarithmically spaced sizes spanning $g=2$-$10^{7}$, at bath gas pressure $p=10^{5}$ Pa and saturation ratios $S=0.1$, $1$, and $10$, with bath gas temperatures of $160$-$280$ K (in $40$ K increments) for water and $500$-$2000$ K (in $500$ K increments) for silver and gold, using the same property correlations, fitted $a_g$, $b_g=b_0g$, and derived $L_g(T)$ as the theoretical calculations.

\subsection{Mean First-Passage (First-Appearance) Time}

To connect the ensemble-averaged survival probability $\Pi_g$ to a cumulative growth timescale, we follow a tagged cluster lineage whose size changes by single-monomer addition and loss.  Considering a cluster initially at size $g$, we denote the time at which it first reaches a target size $G$ by $t_{g\rightarrow G}$, and its mean over repeated size-space trajectories by $\langle t_{g\rightarrow G}\rangle$.  This single-lineage mean first-passage time is also the mean first-arrival (or first-appearance) time at size $G$ for one tagged cluster, and is calculated as:

\begin{equation}
  \left\langle t_{2\rightarrow G}\right\rangle_{\rm surv}
  =
  \sum_{g=2}^{G-1}
  \frac{1}{k_{f,g}[\ce{A_1}]\Pi_g}.
  \label{eq:mfpt_survival}
\end{equation}
Equation (\ref{eq:mfpt_survival}) is the mean first-passage time of the forward-only effective chain.  It omits reversible backtracking in cluster-size space and is therefore not used as a complete nucleation timescale.  However, it is useful to compare this expected "survival-controlled" first-appearance time to those calculated from CNT. For the reversible classical reference, CNT supplies the capillarity-based
work of cluster formation, while a Becker--D\"oring-type birth--death process
supplies the size-space kinetics \cite{Mcgrath_2012,Wedekind_2007}.  We write
\begin{align}
  W_g^{\rm CNT}
  &=
  \theta\left(g^{2/3}-1\right)
  -(g-1)k_{\rm B}T_{\rm gas}\ln S_{\rm m}, \notag\\
  \theta
  &=
  (36\pi)^{1/3}\sigma v_{\rm m}^{2/3},
  \qquad
  \rho_g^{\rm CNT}
  =
  \exp\left(-\frac{W_g^{\rm CNT}}{k_{\rm B}T_{\rm gas}}\right),
  \label{eq:cnt_work_mfpt}
\end{align}
where the statistical weights are normalized by $\rho_1^{\rm CNT}=1$.  The
forward rate is
$\lambda_g^{\rm CNT}=k_{f,g}[\ce{A_1}]$, and the backward rate
$\mu_g^{\rm CNT}$ is fixed by detailed balance,
$\rho_g^{\rm CNT}\lambda_g^{\rm CNT}
=\rho_{g+1}^{\rm CNT}\mu_{g+1}^{\rm CNT}$.  The mean first-passage time obeys
the backward equation
\begin{equation}
  -1
  =
  \lambda_g\left(T_{g+1}^{(G)}-T_g^{(G)}\right)
  +
  \mu_g\left(T_{g-1}^{(G)}-T_g^{(G)}\right),
  \label{eq:mfpt_backward}
\end{equation}
with $T_G^{(G)}=0$ at the absorbing boundary and $\mu_1=0$ at the lower
boundary.  For a lineage initialized as a dimer, its closed-form solution is
\begin{equation}
  \left\langle t_{2\rightarrow G}\right\rangle_{\rm CNT}
  =
  \sum_{g=2}^{G-1}
  \frac{1}{\lambda_g^{\rm CNT}\rho_g^{\rm CNT}}
  \sum_{j=1}^{g}\rho_j^{\rm CNT}.
  \label{eq:mfpt_cnt}
\end{equation}

The ensemble-averaged survival probability $\Pi_g$ contains not only the effect of latent heating, but also the competition between the next monomer arrival and
dissociation from an otherwise equilibrated cluster.  Directly multiplying
$\lambda_g^{\rm CNT}$ by $\Pi_g$ while retaining
$\mu_g^{\rm CNT}$ would therefore count the equilibrium
association--dissociation competition twice.  To isolate the additional
forward penalty associated with post-collision excitation, we evaluate the
same trajectory functional with $L_g=0$, denote the resulting survival
probability by $\Pi_g^{\rm iso}$, and define the operational thermal factor
\begin{equation}
  \varphi_g=\frac{\Pi_g}{\Pi_g^{\rm iso}}.
  \label{eq:thermal_factor}
\end{equation}
The thermal forward-penalty birth--death closure is then defined explicitly
by
\begin{equation}
  \lambda_g^{\rm th}
  =
  \varphi_g\lambda_g^{\rm CNT},
  \qquad
  \mu_g^{\rm th}
  =
  \mu_g^{\rm CNT}.
  \label{eq:thermal_bd_rates}
\end{equation}
The corresponding statistical weights and mean first-passage time are
\begin{align}
  \rho_1^{\rm th}&=1,\qquad
  \rho_g^{\rm th}
  =
  \rho_g^{\rm CNT}
  \prod_{m=1}^{g-1}\varphi_m, \label{eq:thermal_bd_weight}\\
  \left\langle t_{2\rightarrow G}\right\rangle_{\rm th}
  &=
  \sum_{g=2}^{G-1}
  \frac{1}{\lambda_g^{\rm th}\rho_g^{\rm th}}
  \sum_{j=1}^{g}\rho_j^{\rm th}.
  \label{eq:mfpt_thermal}
\end{align}
Because $\varphi_g$ is calculated beginning at $g=2$, the unresolved
monomer--dimer factor is extrapolated as $\varphi_1=\varphi_2$ in the
calculations below.  Products and sums in equations
(\ref{eq:thermal_bd_weight}) and (\ref{eq:mfpt_thermal}) are evaluated in
logarithmic form to avoid numerical overflow.  Equation
(\ref{eq:thermal_bd_rates}) is a deliberately defined kinetic closure; only
the forward rates are modified, so $\rho_g^{\rm th}$ is not asserted to be a
new thermodynamic equilibrium distribution.  Within this closure,
$0<\varphi_g\leq 1$ implies
$\langle t_{2\rightarrow G}\rangle_{\rm th}\geq
\langle t_{2\rightarrow G}\rangle_{\rm CNT}$; a thermal curve above the CNT
reference therefore represents slower first passage, rather than enhanced
nucleation.

\section{Results \& Discussion}
\subsection{Ensemble-Averaged Survival Probabilities}
\begin{figure*}
    \centering
    \includegraphics[width=\textwidth]{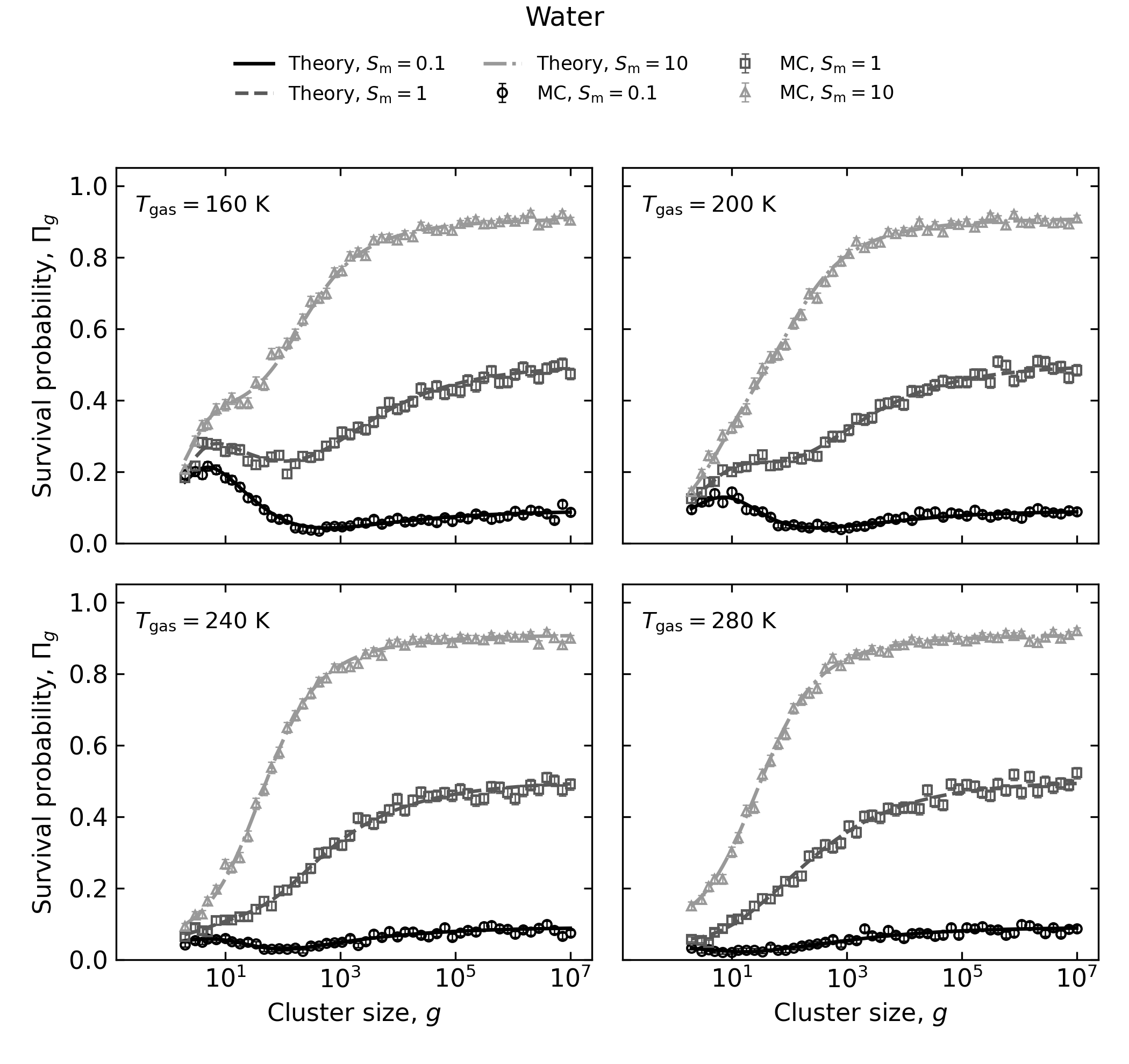}
    \caption{\label{fig:Pi_water} The ensemble-averaged survival probability $\Pi_g$ as a function of cluster monomer number (size $g$) for water, for selected bath gas temperatures and saturation ratios.  Symbols denote Monte Carlo (MC) simulations and lines equation (\ref{eq:central_eqn}) predictions. The bath gas is kept at atmospheric pressure.}
\end{figure*}

\begin{figure*}
    \centering
    \includegraphics[width=\textwidth]{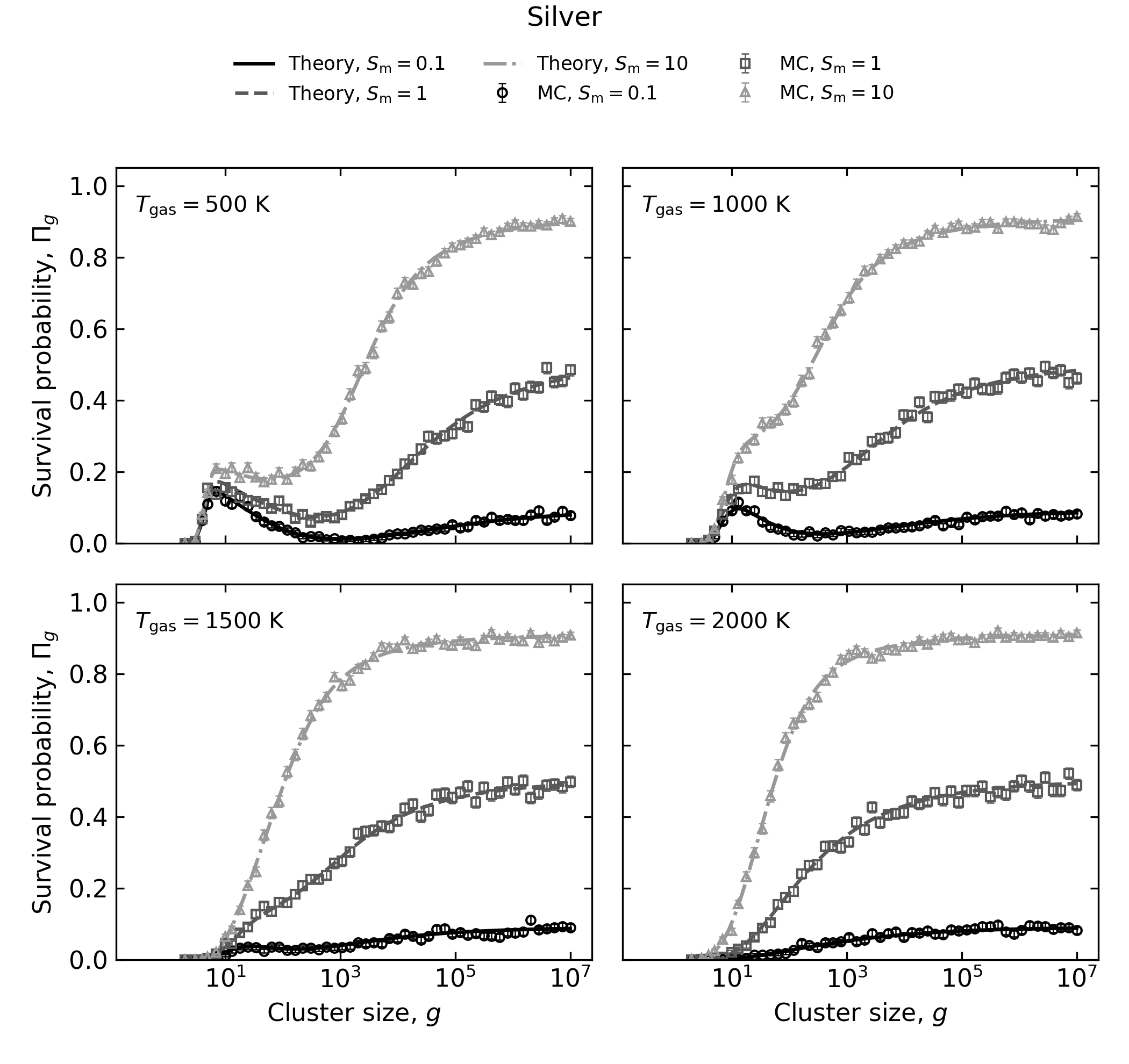}
    \caption{\label{fig:Pi_silver} The ensemble-averaged survival probability $\Pi_g$ as a function of cluster monomer number (size $g$) for silver, for selected bath gas temperatures and saturation ratios.  Symbols denote Monte Carlo (MC) simulations and lines equation (\ref{eq:central_eqn}) predictions. The bath gas is kept at atmospheric pressure.}
\end{figure*}

\begin{figure*}
    \centering
    \includegraphics[width=\textwidth]{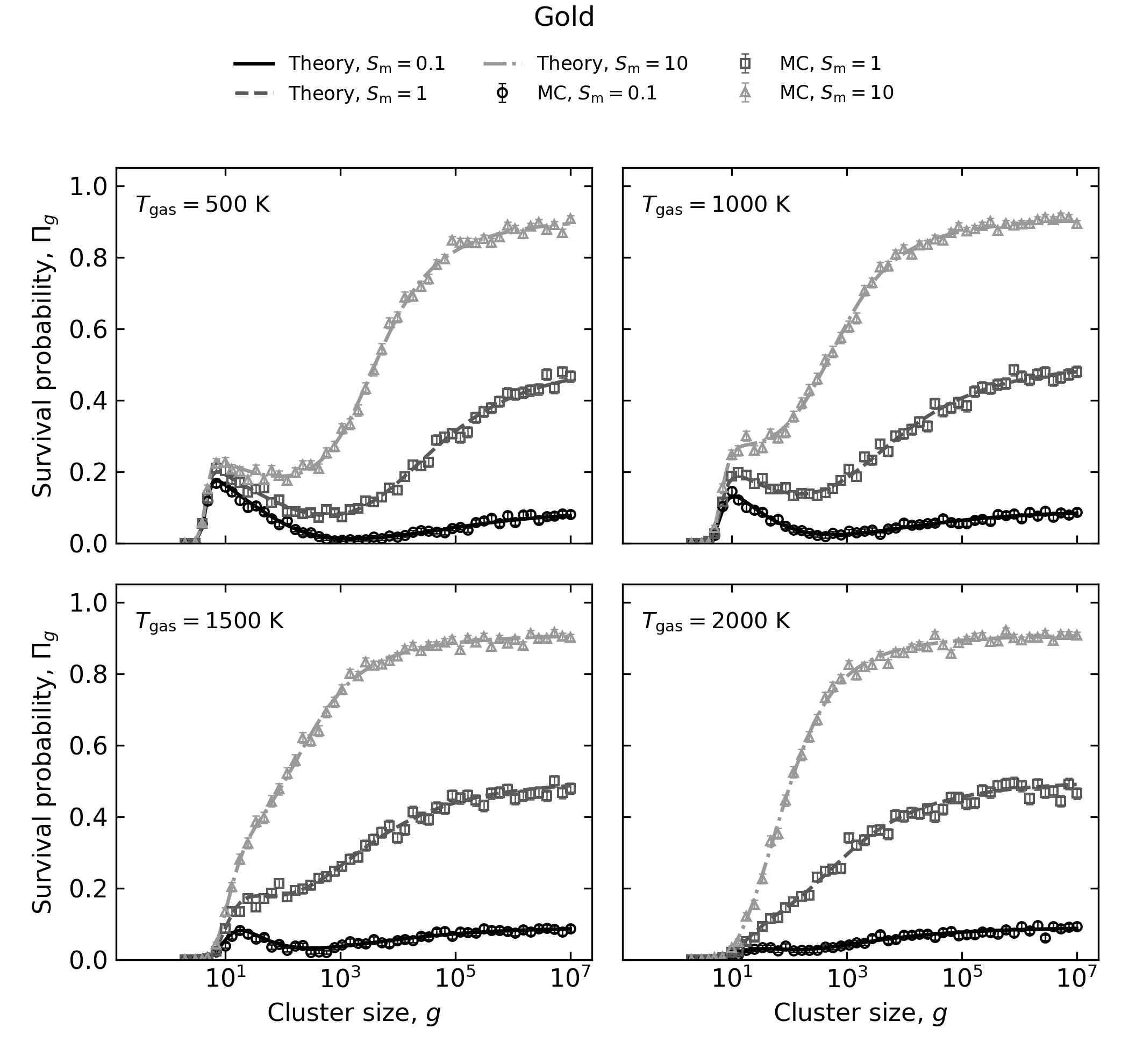}
    \caption{\label{fig:Pi_gold} The ensemble-averaged survival probability $\Pi_g$ as a function of cluster monomer number (size $g$) for gold, for selected bath gas temperatures and saturation ratios.  Symbols denote Monte Carlo (MC) simulations and lines equation (\ref{eq:central_eqn}) predictions. The bath gas is kept at atmospheric pressure.}
\end{figure*}

\begin{figure*}
    \centering
    \includegraphics[width=\textwidth]{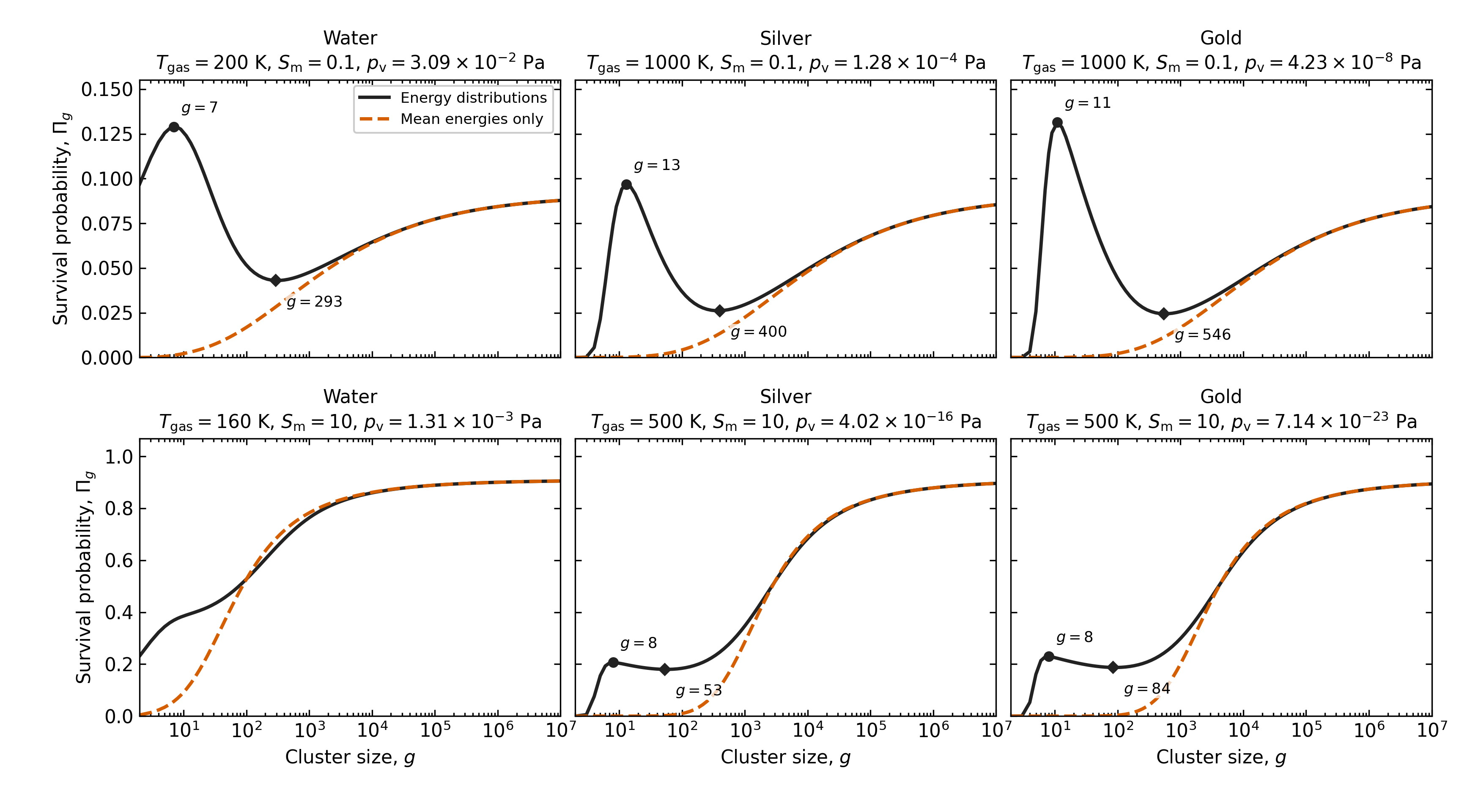}
    \caption{\label{fig:energy_mean} The ensemble-averaged survival probability $\Pi_g$ as a function of cluster monomer number (size $g$) predicted using complete energy distributions for $f_{\rm ex(E_{\rm ex})}$ and $f_{\rm eq}(E_{\rm eq})$ (solid lines) and using mean values for $f_{\rm ex}(E_{\rm ex})$ and $f_{\rm eq}(E_{\rm eq})$.}
\end{figure*}

Growing cluster survival based on competing thermal relaxation and dissociation until a subsequent monomer collision yields the ensemble-averaged survival probability as the main metric to assess the influence of thermal excitation on cluster growth. Figures \ref{fig:Pi_water}, \ref{fig:Pi_silver}, and \ref{fig:Pi_gold} plot $\Pi_g$ versus cluster size for water, silver, and gold, respectively, comparing equation (\ref{eq:central_eqn}) predictions to Monte Carlo simulations, for variable saturation ratios and temperatures. Immediately evident is the strong agreement between predictions and Monte Carlo simulations, validating the QSSA and thermal trajectory approach to determine survival.  We can then proceed to examine $\Pi_g$ versus $g$ curves, which contain several features consistent across three examined cluster types and temperatures.  As $g$ increases, $\Pi_g$ values approach an asymptotic limit which increases with increasing saturation ratio.  This is expected; as size increases latent heating effects eventually become negligible $ T_{\rm l}(t) \rightarrow T_{\rm gas}$, and $\Pi_g$ simplifies to

\begin{equation}
    \lim_{g\rightarrow\infty}\Pi_g
    =
    \lim_{g\rightarrow\infty}\Pi_g^{\rm iso}
    =
    \frac{S_{\rm m}}{S_{\rm m}+\exp\left[
    \frac{2\sigma v_{\rm m}}
    {k_{\rm B}T_{\rm gas}a_{g+1}}
    \right]},
    \label{eq:Pi_large_g_asymptote}
\end{equation}

\noindent which further simplifies to $\lim_{g\rightarrow\infty}\Pi_g=\frac{S_{\rm m}}{S_{\rm m}+1}$ when the Kelvin effect is negligible.  Equation (\ref{eq:Pi_large_g_asymptote}) would also directly result from a simplified reaction scheme, wherein monomer growth event yields cluster $\ce{A}_{g}+\ce{A_1}\rightarrow \ce{A}_{g+1}$ which either dissociates or grows further, and the concentration $[\ce{A}_{g+1}]$ is assumed to be in steady state. Prior to the asymptote, $\Pi_g$ largely decreases with decreasing $g$, reaching a minimum below $g=10$.  This demonstrates the thermal excitation barrier is most pronounced for the smallest clusters, which again is expected as the Kelvin effect leads greatly enhanced dissociation rates in this size which are further exacerbated by thermal excitation.  Interestingly, however, in some instances we observe a local maximum in $\Pi_g$ in the $g=10^1-10^2$ range.  This maximum is more pronounced at at lower temperatures, and lower saturation ratios. Unlike the large size and small size limiting behavior is not immediately apparent why such local maxima manifest.  We find they arise due to the shape of the energy distributions $f_{\rm ex}$ and $f_{\rm eq}$. In figure \ref{fig:energy_mean} we plot $\Pi_g$ comparing predictions using complete energy distributions to use of mean energies only, i.e. $f_{\rm ex}(E_{\rm ex})=\delta(E_{\rm ex}-[\frac{\nu}{2}(1+a_g)gk_BT_{gas}+b_{g}+L_g])$ and $f_{\rm eq}(E_{\rm eq})=\delta(E_{\rm eq}-[\frac{\nu}{2}(1+a_{g+1})(g+1)k_BT_{gas}+b_{g+1}])$. Without including full energy distribution functions, local maxima do not appear. Although the Kelvin effect is important at small sizes, its influence via the inverse of the characteristic dissociation time scales as $\tau_d^{-1}\propto\exp(\frac{B_1}{k_BT}(g+1)^{-1/3})$, which decreases rapidly with increasing size.  Meanwhile the relative with of the energy distributions scales with $g^{-1/2}$, which is a less steep change with size, and leads to broad energy distributions at small size.  There is thus a less energetic (i.e. colder) population of small size clusters which can survive the growth process, but whose relative fraction decreases with increasing size.  In the $g=10^1-10^2$ range, for certain material, temperature, and saturation ratio combinations, the interplay between these two size-dependent effects leads to the local survival probability maximum.

\begin{figure*}
    \centering
    \includegraphics[width=\textwidth]{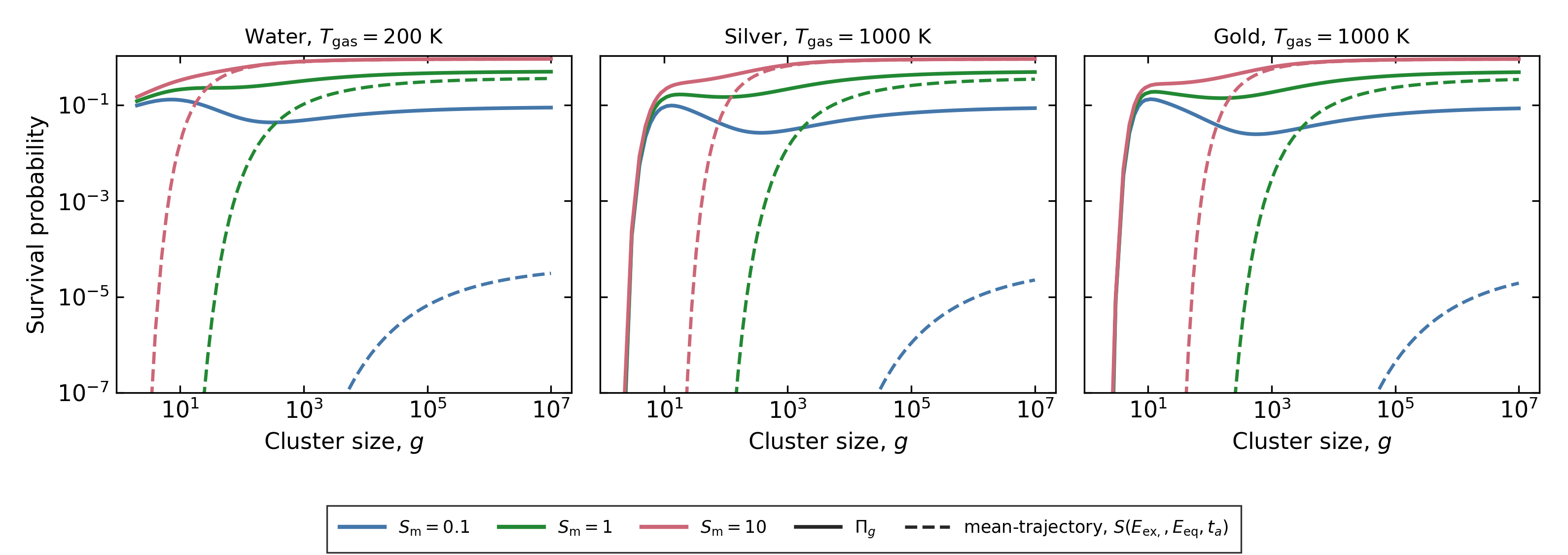}
    \caption{\label{fig:ensemble_vs_single} $\Pi_g$ versus clusters size $g$ solid lines in comparison to $S(E_{\rm ex}, E_{\rm eq},t_a)$} using mean values to define the thermal trajectory.
\end{figure*}

The importance of marginalizing over all thermal trajectories to compute the ensemble-averaged survival probability is further shown in figure \ref{fig:ensemble_vs_single} in comparison to the single thermal trajectory survival probability $S(E_{\rm ex}, E_{\rm eq},t_a)$ using mean values to define the thermal trajectory. At low saturation ratios, and importantly for smaller clusters, the mean thermal trajectory survival probability deviates significantly from the ensemble-averaged survival probability.  Deviations can extend beyond several orders of magnitude with $\Pi_g\geq S(E_{\rm ex}, E_{\rm eq},t_a)$ arising because of the low energy tail of energy distributions.

\subsection{Surviving Cluster Populations}
\begin{figure*}
    \centering
    \includegraphics[width=\textwidth]{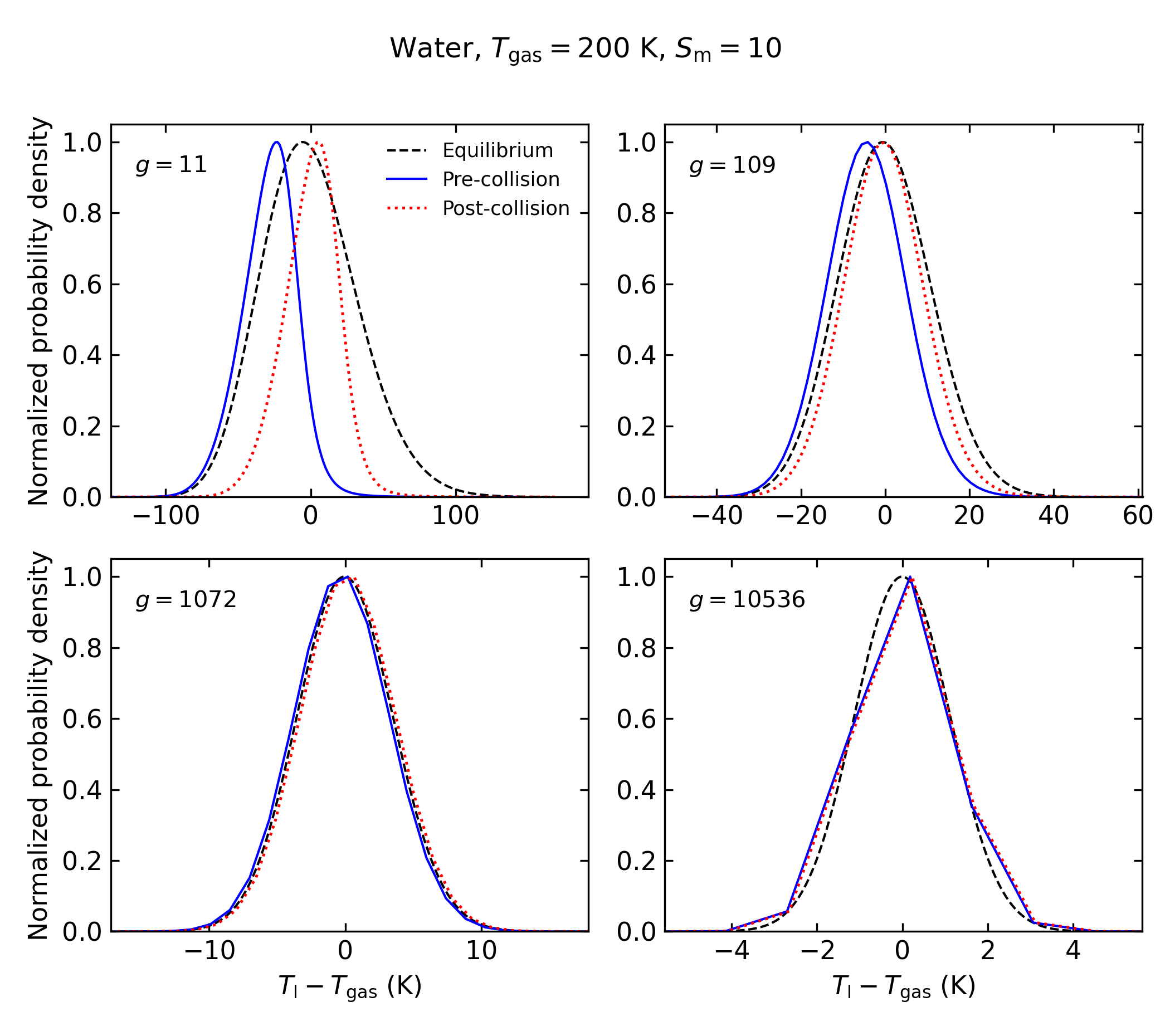}
    \caption{\label{fig:surv_dist_water} The equilibrium energy distribution $f_{\rm eq}(E_{\rm eq})$, the initial energy distribution of clusters which survive (pre-collision) resulting from Monte Carlo simulations, and the energy of distribution of surviving clusters (post-collision), which is shifted by $L_g(T)$ for water clusters.}
\end{figure*}

\begin{figure*}
    \centering
    \includegraphics[width=\textwidth]{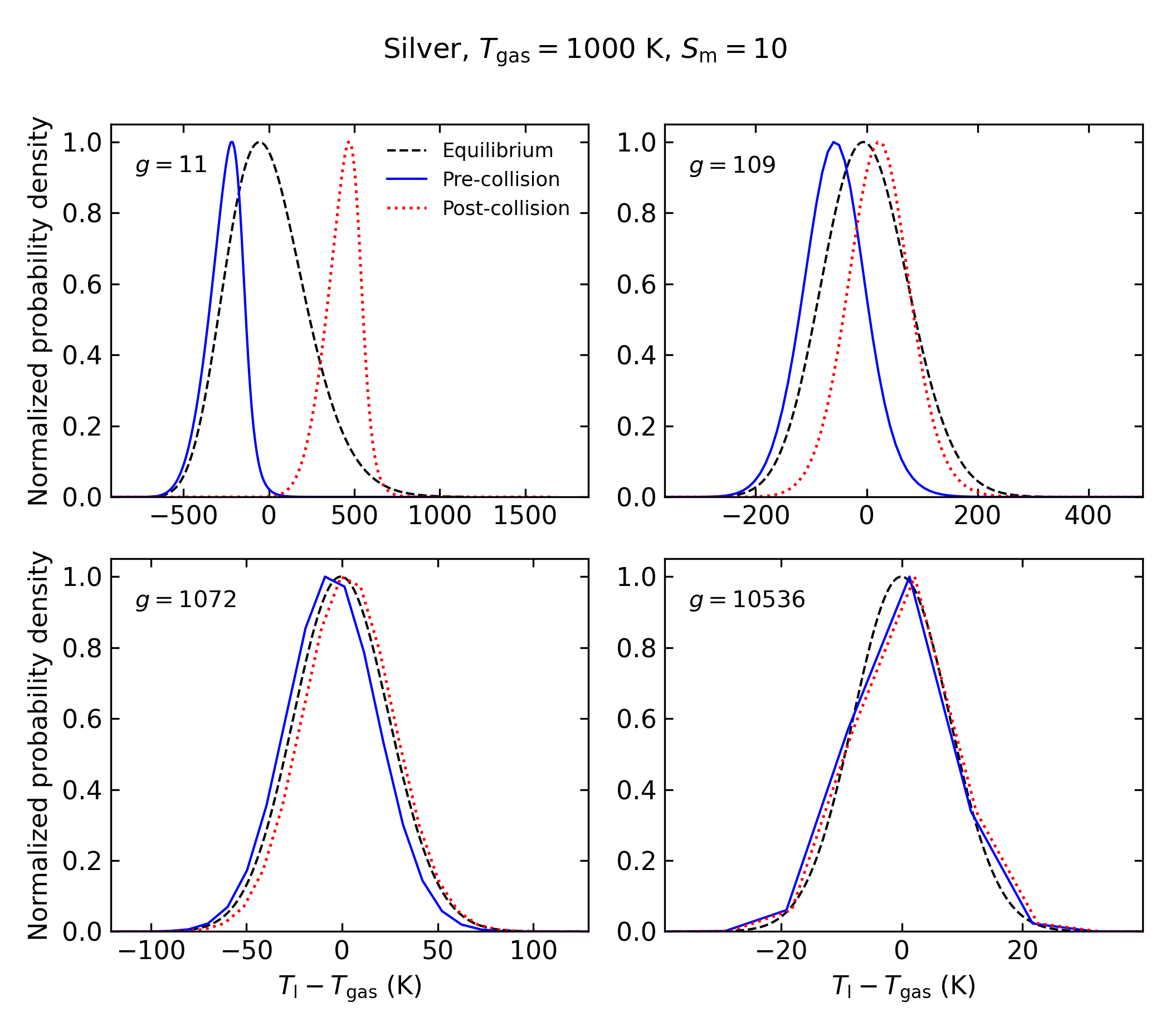}
    \caption{\label{fig:surv_dist_silver} The equilibrium energy distribution $f_{\rm eq}(E_{\rm eq})$, the initial energy distribution of clusters which survive (pre-collision) resulting from Monte Carlo simulations, and the energy of distribution of surviving clusters (post-collision), which is shifted by $L_g(T)$ for silver clusters.}
\end{figure*}

\begin{figure*}
    \centering
    \includegraphics[width=\textwidth]{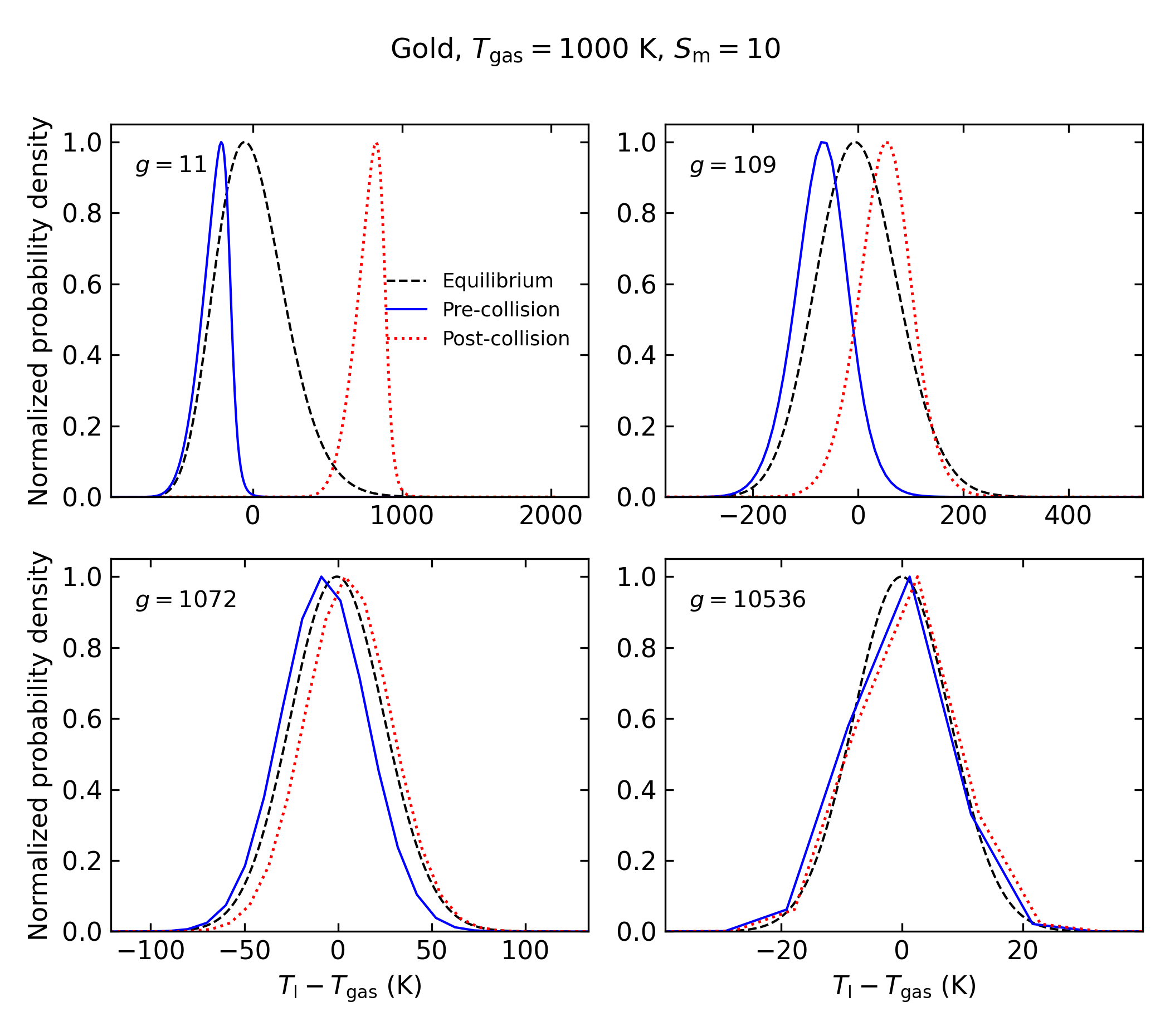}
    \caption{\label{fig:surv_dist_gold} The equilibrium energy distribution $f_{\rm eq}(E_{\rm eq})$, the initial energy distribution of clusters which survive (pre-collision) resulting from Monte Carlo simulations, and the energy of distribution of surviving clusters (post-collision), which is shifted by $L_g(T)$ for gold clusters.}
\end{figure*}

Observation of local maxima arising due to the span of the energy distribution suggests that clusters which do survive the growth process are biased in initial energy, also identified in full MD simulations of nucleation \cite{Toxvaerd_2015}.  To examine this, in figures \ref{fig:surv_dist_water}, \ref{fig:surv_dist_silver}, and \ref{fig:surv_dist_gold}, we plot the equilibrium energy distributions for clusters (size $g+1$) as well as the Monte Carlo simulation-inferred pre- and post-collision energy distributions for clusters which do survive the thermal excitation process. Distributions are expressed in terms of temperature difference between the cluster $T_{l}$ and the equilibrium gas temperature $T_{gas}$.  Across all test cases results show that for $g=10^2$ and smaller, the clusters able to survive the growth process are predominantly colder than the bath gas by $10^1$ to $10^2$ K in effective temperature.  Upon collision, their energy distributions are shifted by $L_g$, where with the exception of the smallest silver and gold clusters examined, many clusters still have energies at or below the mean energy based upon the bath gas temperature. This highlights that thermal excitation strongly increases dissociation propensity and biases growth towards the lower energy tails of distributions.  Despite extrapolating the thermal properties of these clusters from simulation, this finding likely holds even with more detailed models of cluster thermal properties, as it follows from the span of the energy distribution. Solution to complete two-dimensional size-energy population balances is necessary to better establish cluster energy distribution functions before collision events as they may deviate from equilibrium distributions. We are therefore not able to comment on whether clusters in nucleating systems at variable sizes are hotter or colder than the surrounding gas, which has been a central question in prior work \cite{Feder1966, Valtteri_2022,Barrett_2008}.  However, the finding that lower energy clusters preferentially survive will still hold irrespective of assumed or modeled shape for excited state distributions.  In full two-dimensional distribution solutions, including not only monomer uptake but also cluster-cluster coagulational growth, Chen et al \cite{CHEN2024} observed a population of clusters appreciably colder than the bath gas, although the mean temperature across multiple clusters was higher. Our results confirm that this colder population is more likely to control growth to larger sizes.  In the large size limit (approximately $g>10^3$), for all simulation conditions we find a diminished influence of thermal excitation and non-isothermal effects in general. 

\subsection{Cumulative Growth Times in Cluster-Size Space}

\begin{figure*}
    \centering
    \includegraphics[width=\textwidth]{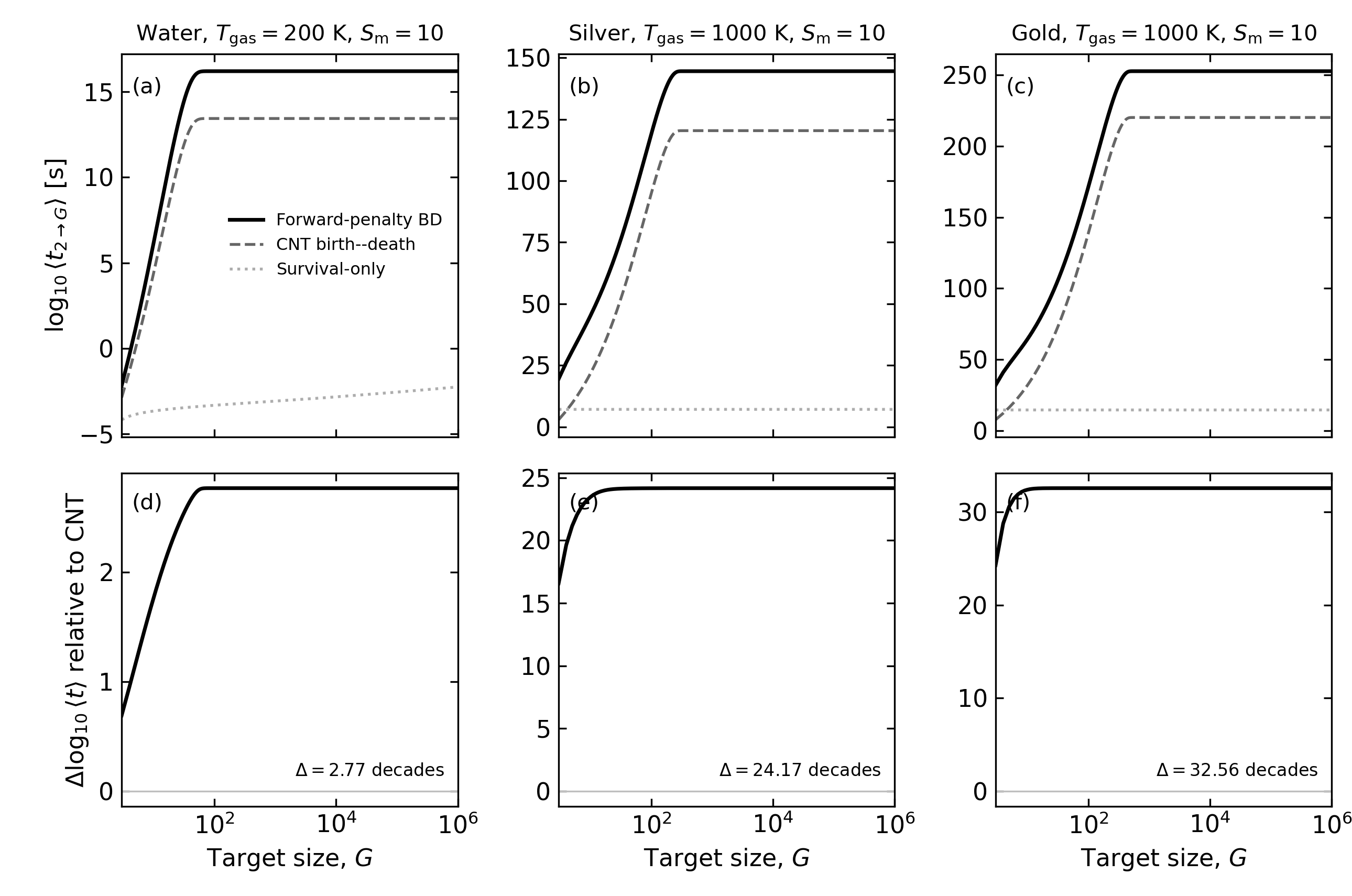}
    \caption{\label{fig:first_passage}Mean first-passage times for a tagged cluster lineage initialized at $g=2$ and absorbed upon first reaching target size $G$.  Panels (a)--(c) compare the thermal forward-penalty birth--death closure (solid), the CNT-parameterized birth--death baseline
    (dashed), and the forward-only survival-weighted traversal time from equation (\ref{eq:mfpt_survival}) (dotted).  Panels (d)--(f) show $\Delta\log_{10}\langle t\rangle
    =\log_{10}\langle t\rangle_{\rm th}
    -\log_{10}\langle t\rangle_{\rm CNT}$.
    All calculations use $p=1.0\times10^5$ Pa and $S_{\rm m}=10$; the temperatures are $200$ K for water and $1000$ K for silver and gold.
    The forward-only curve omits size-space backtracking and is included as a
    diagnostic rather than as a complete nucleation timescale.}
\end{figure*}

In figure \ref{fig:first_passage}(a-c) we plot the $\log_{10}$ of the mean first passage times considering survival only, which would be equivalent to collision-limited growth plus a survival probability for single thermal excitation events (equation (\ref{eq:mfpt_survival})), the mean first passage times based on CNT only (isothermal, equation (\ref{eq:mfpt_cnt}) ), and the mean first passage time considering $\varphi_g$ in defining mean first passage times (equation (\ref{eq:mfpt_thermal}).  In figure \ref{fig:first_passage}(d-f) we plot the $\log_{10}$ difference between isothermal CNT and non-isothermal mean size passage time results. Such plots ioslate the cumulative influence of the
single-step survival probabilities from the reversible first-passage problem. Using the smooth size-dependent caloric model extrapolated to small $g$, the thermal forward-penalty closure at $G=10^6$ increases
$\log_{10}\langle t_{2\rightarrow G}\rangle$ relative to the
CNT-parameterized birth--death baseline by $2.77$ decades for water,
$24.17$ decades for silver, and $32.56$ decades for gold.  In each case the
difference approaches a plateau at small $G$, showing that the additional
thermal penalty is accumulated before the later CNT-barrier contribution and
is not generated by continued growth at large $G$.  Specifically, $95\%$ of
the final offset is accumulated by $G=39$, $8$, and $6$ for water, silver, and gold, respectively.  These sizes lie below the MD fitting ranges ($g\geq107$ for water and $g\geq3043$ for the metals), so the numerical offsets are sensitive to the small-cluster extrapolation, the $\varphi_1=\varphi_2$ lower-boundary convention, and the forward-penalty closure itself.  They should therefore be interpreted as model-sensitivity results, not as universal corrections to CNT or direct measurements of a population nucleation rate. Nonetheless, this comparison confirms that the multiplicative effects of thermal-excitation induced dissociation can be substantial on particle formation, and in many instances may contribute to observed differences between measured nucleation rates and CNT \cite{Wagner_1981,Strey_1986,Brus_2009,Wyslouzil_2016}.  Our model  contains an explicit pressure dependency consistent in form with recent measurements \cite{Campagna_2020}. The need for clusters to survive thermal-excitation is also consistent with the recent work of Li, Signorell, and co-workers \cite{Li_2021,Feusi_2024,Choudhury_2026}, who find that  clusters formed in heterogeneous mixtures of water or organic species with carbon dioxide grow faster likely because of carbon dioxide clustering and dissociation, which can reduce the thermal excitation penalty without effecting growth of the less volatile species.  However, direct comparison of the present approach to experimental measurements remains complicated by invoking the typical assumptions of CNT for cluster dissociation properties, particular for the $g\leq10^1$ size range, as well as the need for more accurate thermal accommodation coefficients \cite{Sipkens_2018}.

\section{Conclusion}
We developed a survival-limited description of gas-phase cluster growth by
monomer addition in which a collision does not necessarily constitute a
successful growth event. Instead, successful growth therefore requires that the newly formed cluster survive a post-association interval until the next growth event, during which time it is thermally-excited and prone to dissociation. Starting from a discrete sequence of energetic relaxation states, we derived a continuous-energy survival probability for an individual thermal trajectory of a cluster. We then marginalized this trajectory-level probability over distributions of post-association excitation energy, equilibrium energy, and next-monomer arrival time to obtain the ensemble-averaged survival probability $\Pi_g$.
Event-based Monte Carlo simulations, which directly sampled the competing
arrival and dissociation events without evaluating the analytical survival
integral or invoking the quasi-steady-state derivation, agreed closely with
the trajectory-functional predictions for water, silver, and gold clusters.  Based on this model and its validation, we make the following concluding remarks:
\begin{itemize}
    \item The calculated survival probabilities exhibit a strong and non-monotonic dependence on cluster size. Survival is generally lowest for the smallest clusters because latent heating and curvature-enhanced dissociation act simultaneously. At intermediate sizes, local maxima can arise from competition between the rapidly diminishing Kelvin contribution and the more slowly narrowing cluster-energy distribution. These maxima disappear when the full energy distributions are replaced by their mean values, demonstrating that cluster-population survival cannot generally be inferred from a single mean thermal trajectory.
    \item Marginalization over the full trajectory distribution is particularly important for small clusters and low saturation ratios. Under these conditions, the ensemble-averaged survival probability can exceed the survival probability calculated from mean trajectory parameters by several orders of magnitude. Growth is consequently biased toward clusters drawn from the low-energy tail of the pre-collision population. The clusters that advance to larger sizes are therefore not a thermally representative sample of the population from which they originated. 
    \item At sufficiently large sizes, latent-heating and curvature effects diminish, and the survival probability approaches the ordinary competition between equilibrium monomer dissociation and arrival of the next monomer. This large-size behavior also clarifies why $\Pi_g$ itself should not be applied directly as a correction to a reversible classical nucleation model. The equilibrium association--dissociation competition represented in $\Pi_g$ is already present in the classical birth--death description. We therefore introduced the thermal correction factor $\varphi_g$, defined as the ratio of the complete survival probability to an isothermal-reference survival probability calculated using the same trajectory functional with latent heating removed. This isolates the additional forward kinetic penalty associated specifically with post-association thermal excitation. Although the thermal correction at any individual cluster size may appear moderate, its effect is accumulated across a reversible sequence of monomer-addition and monomer-loss events.
    \item The present calculations intentionally use smooth size-dependent caloric relationships and bulk-inspired models for dissociation and thermal relaxation. These closures can be replaced without changing the structure of the survival framework. Future work should incorporate cluster-resolved binding energies, heat capacities, latent heats, structures, and dissociation rates obtained from atomistic simulation, electronic-structure calculations, or experiment. Such refinement will be especially important for magic-number clusters \cite{Girshick_2009}, for which enhanced structural or electronic stability may produce abrupt size-dependent changes that cannot be represented by a smooth continuum correction.
    \item The same competing-risk construction can be extended to heterogeneous growth. For example, a volatile precursor may temporarily bind to the surface of a multicomponent cluster and subsequently cool, desorb, or undergo a surface reaction that converts it into a less volatile incorporated species. In such a system, successful growth would correspond to reaction before desorption, rather than simply survival until another monomer arrives. Reaction and desorption could be represented as competing, energy-dependent hazards along the post-collision thermal trajectory.
    \item The trajectory space can also be expanded for non-equilibrium environments such as plasmas \cite{LI_2025} and even flames \cite{WANG_2011}. In these environments, collision outcomes may depend explicitly on relative speed, impact parameter, cluster charge, monomer charge, internal state, electronic excitation, and collision orientation. These variables can be included directly in the trajectory description, with excitation energies and competing hazards conditioned on the collision parameters.
    \item More generally, the trajectory-functional route applied here separates the outcome associated with one specified post-collision trajectory and the statistical distribution of trajectories present in the physical system. Improvements to either component can be introduced independently. The thermal path may be deterministic or stochastic, the relevant hazards may be obtained from continuum theory, molecular simulation, or experiment, and the trajectory distribution may be equilibrium, non-equilibrium, correlated, or empirically inferred. The factorized trajectory distribution used in the present calculations is therefore a closure rather than a fundamental requirement.
\end{itemize}

\section*{Supplementary Material}
See the Supplementary Material for the complete size-dependent caloric-property fits, comparison with bulk-property constants, model-selection diagnostics, and structural analysis supporting the fitting range used for the metal clusters.

\section*{Acknowledgments}
T.T. was supported by JSPS KAKENHI under Grant No. 24K17542.  C.J.H. acknowledges support from the James J. Ryan Professorship from the University of Minnesota College of Science \& Engineering and Department of Mechanical Engineering. Portions of computations were carried out using the computer resource offered under the category of Trial Use Projects by the Research Institute for Information Technology, Kyushu University.

\section*{Data Availability Statement}
The data that support the findings of this study are available within the article.

\section*{Author Declarations}
The authors have no conflicts-of-interest to disclose related to this work.  The authors utilized large language models in the proofreading of this manuscript, checking equations for consistency, and to write out longer equations in LATEX.

\bibliography{JCP.bib}





\end{document}


\raggedbottom

\begin{center}
{\large\bfseries Supplementary Material for\\[0.3em]
``Post-Collision Thermal Excitation and Survival-Limited Cluster Growth in the Gas Phase''\par}
\vspace{1em}
Tomoya Tamadate$^{1}$ and Christopher J. Hogan Jr.$^{2}$\\[0.4em]
{\small
$^{1}$Faculty of Frontier Engineering, Institute of Science and Engineering, Kanazawa University\\
$^{2}$Department of Mechanical Engineering, University of Minnesota
}
\end{center}
\vspace{1em}

\section{Degrees of freedom in the latent-heat relation}

Equation (18) of the main text extends the monatomic condensation relation of Yang et al. \cite{Yang_2019} to a rigid molecular monomer. Let $\nu$ be the number of unconstrained kinetic degrees of freedom per monomer. After removing the three translational degrees of freedom of the cluster center of mass, the equilibrium mean kinetic energy of an $n$-mer is
\begin{equation}
\langle K_n\rangle=\frac{\nu n-3}{2}RT.
\end{equation}
For the linear caloric relation $\langle U_n\rangle=a_n\langle K_n\rangle+b_n$, the difference between the equilibrium internal energies of the two reactants and the product in $\ce A_g+\ce A_1\rightarrow \ce A_{g+1}$ is supplemented by the collision-weighted relative translational energy retained by the product. Under the same spherical-collision and rotational-energy approximation used by Yang et al., this contribution is
\begin{equation}
Q_{\rm coll}=2RT\left[1-\frac{5g}{4(g+1)^2}\right].
\end{equation}
Combining these terms gives
\begin{align}
L_g(T)={}&b_g+b_1-b_{g+1}+RT\left\{\frac{1}{2}\right.\notag\\
&+\frac{1}{2}\left[(\nu g-3)a_g+(\nu-3)a_1\right.\notag\\
&\left.\left.-\{\nu(g+1)-3\}a_{g+1}\right]
-\frac{5g}{2(g+1)^2}\right\}.
\end{align}
For monatomic silver and gold, $\nu=3$, and this expression is exactly equation (11a) of Yang et al. For the rigid nonlinear TIP3P water molecule, $\nu=6$. SHAKE removes intramolecular vibrational degrees of freedom, while the potential energy of an isolated constrained monomer is independent of its rotational kinetic energy; hence $a_1=0$. Equation (18) then follows for all three materials. Application to a flexible molecular model would require including the active vibrational degrees of freedom and, where applicable, a nonzero monomer caloric slope. The water form is therefore a model extension under these assumptions, rather than a result explicitly validated for molecular clusters in the cited monatomic gold study.

\section{Size-dependent caloric-property model}

The caloric-curve slope at each cluster size was obtained from a linear regression of the molecular-dynamics (MD) block averages to $U=a_gK+b_g$. For water, the slope data cover $107\leq g\leq140091$. For silver and gold, a common $200$--$500$ K window was used so that every size was compared over the same low-temperature branch and no regression crossed the melting transition. The leading finite-size model was
\begin{equation}
a_g=a_{\infty}+A g^{-1/3}.
\end{equation}
The $g^{-1/3}$ term is the leading surface-fraction correction for a compact three-dimensional cluster. The fitted parameters were $(a_{\infty},A)=(2.2635,3.1653)$ for water, $(1.0981,0.4360)$ for silver, and $(1.0810,0.4604)$ for gold.

\begin{figure}[H]
    \centering
    \includegraphics[width=0.92\linewidth]{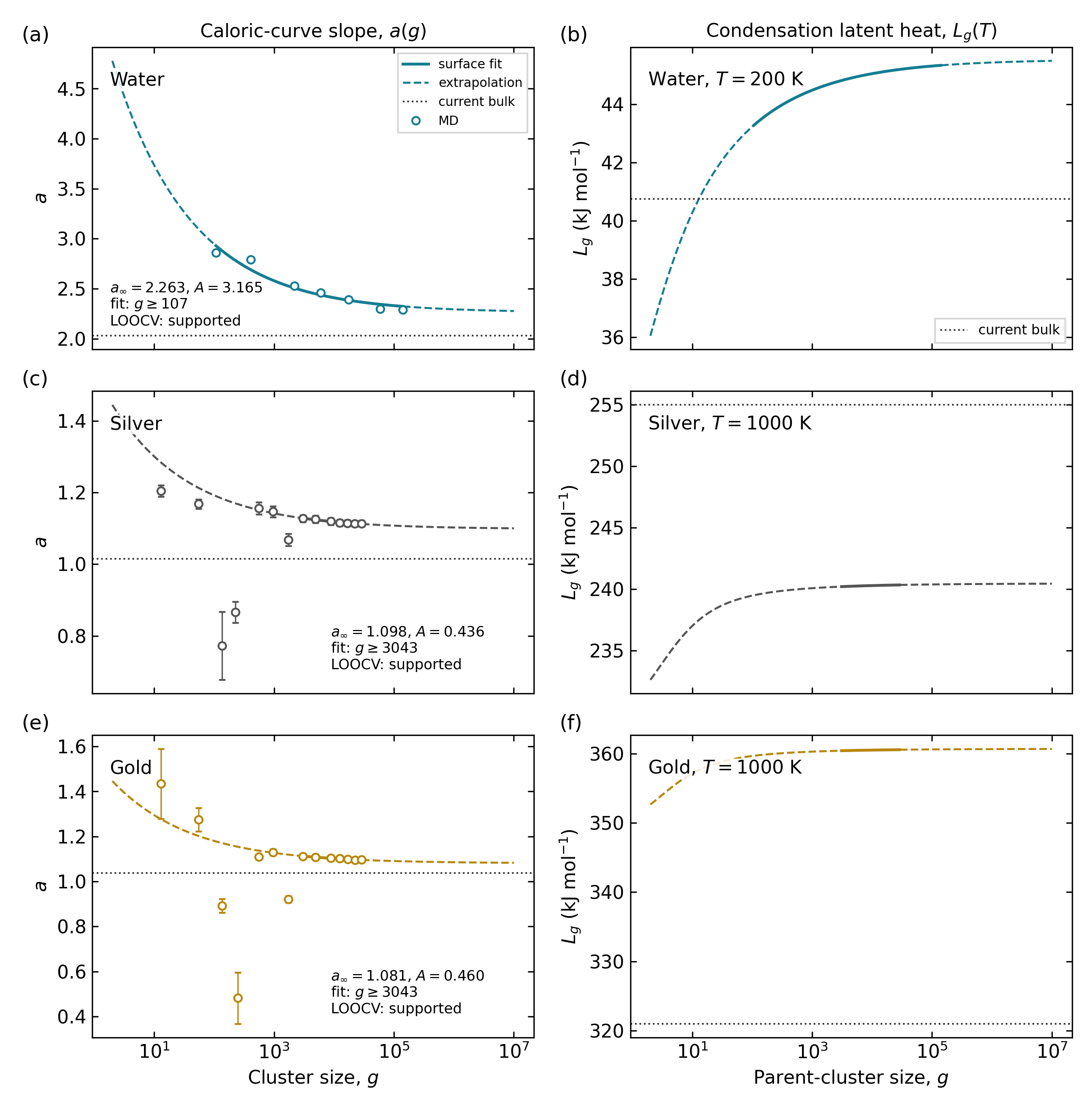}
    \caption{Size-dependent thermal-property modeling. (a), (c), and (e) are the MD caloric slopes for water, silver, and gold, respectively, together with $a_g=a_{\infty}+A g^{-1/3}$. Open symbols show all available MD estimates. Solid curves identify the fitted intervals; dashed curves are extrapolations, and dotted horizontal lines are values inferred from tabulated bulk heat capacities. The metal fits use only $g\geq3043$. (b), (d), and (f) are the latent heat $L_g(T)$ calculated from the fitted caloric model for the indicated temperatures. Dotted lines show tabulated bulk latent heats.}
    \label{fig:supp_thermo}
\end{figure}

Figure~\ref{fig:supp_thermo} shows all MD slope estimates, including metal sizes not used to determine the production fit. Solid curve segments denote the MD-constrained fitting interval; dashed segments are extrapolations. The intercept was represented as $b_g=b_0g$ after the fitted per-monomer intercept approached a material-dependent plateau. The latent heat was not fitted as a second independent function. Instead, $L_g(T)$ was evaluated from the caloric energy balance using $a_g$, $a_{g+1}$, and $b_g$. The right column of Figure~\ref{fig:supp_thermo} compares the resulting size-dependent latent heat with the tabulated bulk value used in the earlier constant-property calculation.

\section{Selection of the size-correction form}

\begin{figure}[H]
    \centering
    \includegraphics[width=0.98\linewidth]{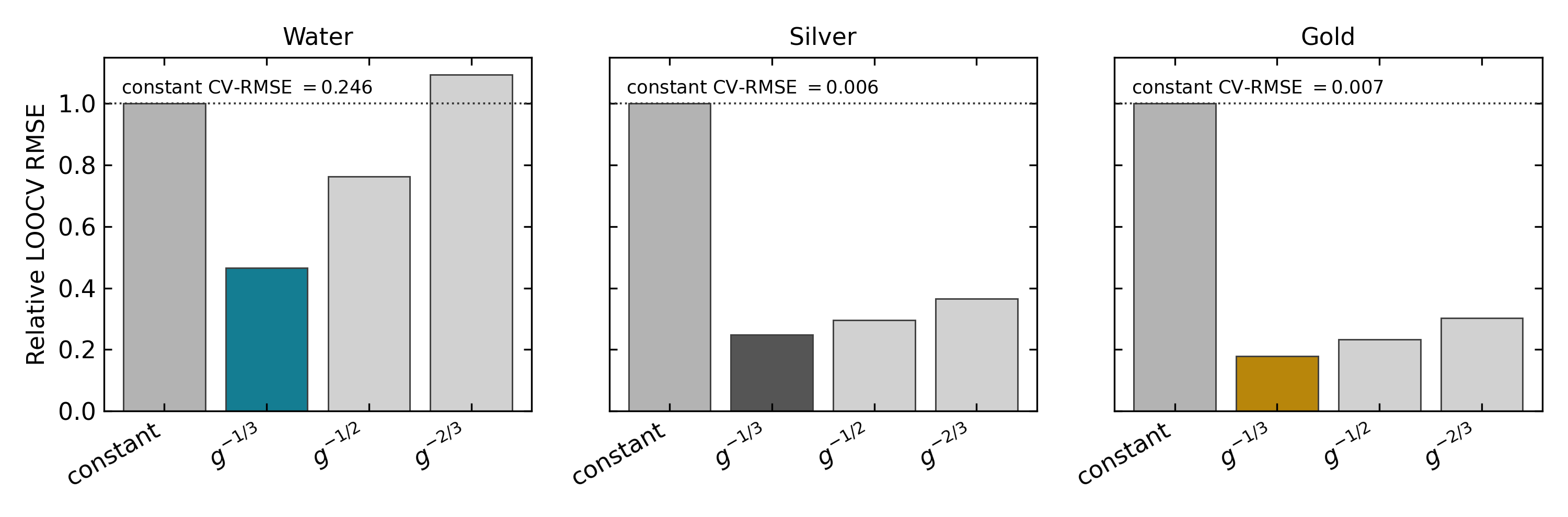}
    \caption{LOOCV comparison of the constant model and three candidate finite-size corrections. Values are normalized by the LOOCV RMSE of the constant model for each material; smaller values indicate better out-of-sample prediction within the stated fitting interval.}
    \label{fig:supp_validation}
\end{figure}

We compared the constant model with $g^{-1/3}$, $g^{-1/2}$, and $g^{-2/3}$ corrections using leave-one-out cross-validation (LOOCV) over the production fitting interval. Figure~\ref{fig:supp_validation} reports each LOOCV root-mean-square error normalized by that of the constant model. The $g^{-1/3}$ form has the lowest prediction error for all three materials. Its relative LOOCV errors are 0.466 for water, 0.248 for silver, and 0.179 for gold. This comparison supports using the leading surface-fraction correction without introducing a higher-parameter expansion that is not warranted by the number of available large-cluster points.

\section{Exclusion of Small Metal Clusters}

The caloric curve parameter fitting we apply assumes a compact particle with a progressively bulk-like interior and a caloric slope that changes smoothly with size. The small silver and gold clusters do not satisfy this asymptotic description. As shown in Figure~\ref{fig:supp_thermo}(c,e), their low-temperature slopes are strongly non-monotonic: several points fall well below the smooth large-size branch, while others lie above it. These deviations are substantially larger than the regression uncertainties and occur for both materials, so including them would force a single smooth surface correction to reproduce discrete, structure-specific behavior.

\begin{figure}[H]
    \centering
    \includegraphics[width=0.94\linewidth]{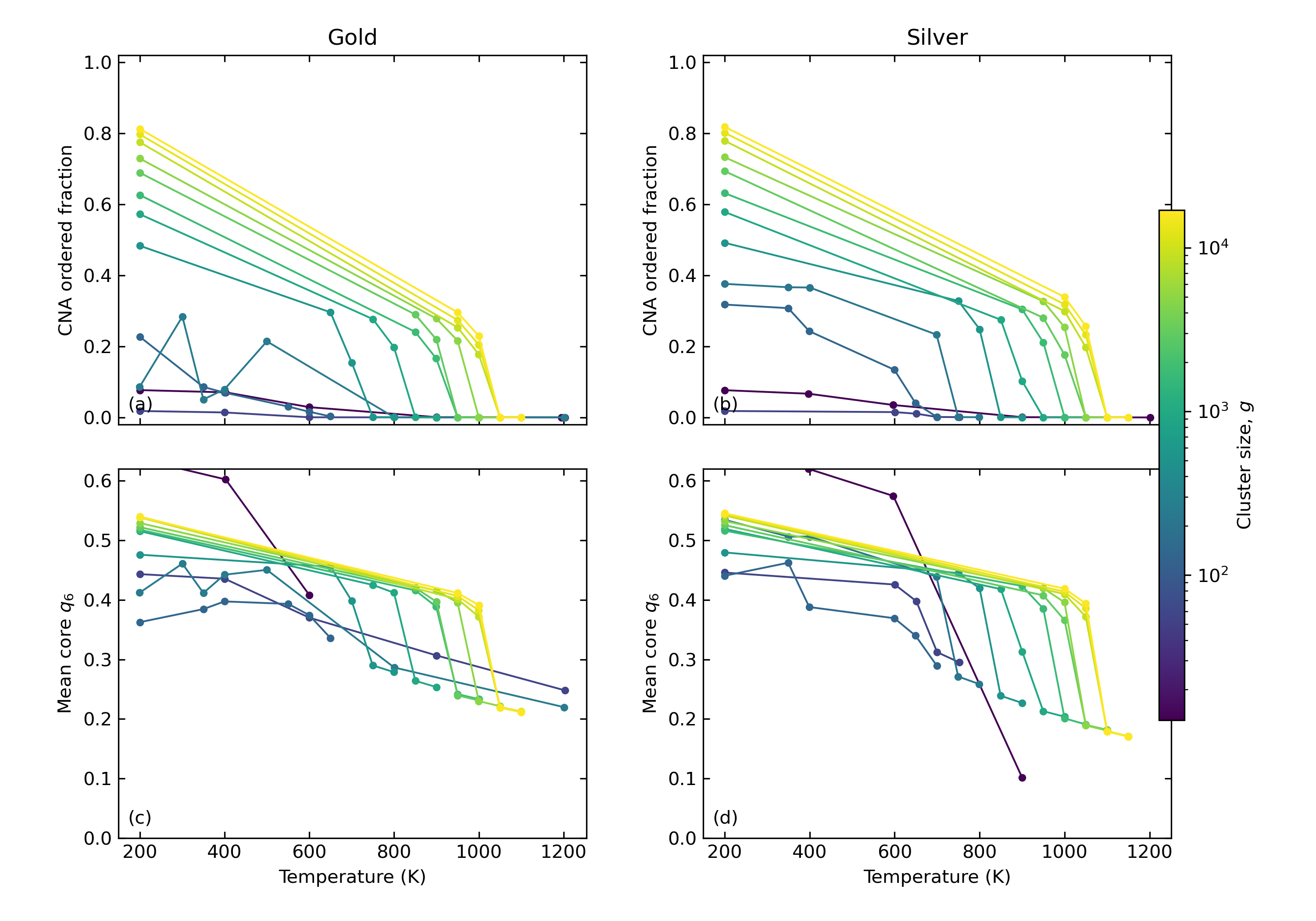}
    \caption{Structural diagnostics from the full metal-cluster scan. (a,b) Fraction of atoms classified as FCC, HCP, BCC, or icosahedral by common-neighbor analysis. (c,d) Mean core-local Steinhardt $q_6$, where the core comprises atoms with coordination number at least 11. Curves are colored by cluster size $g$. The smallest clusters have a small or absent bulk-like core and exhibit a less regular temperature dependence than larger clusters.}
    \label{fig:supp_structure}
\end{figure}

To examine this cluster-size dependent behavior further, during the MD sampling interval, common-neighbor-analysis ordered fractions and core-local Steinhardt $q_6$ values were averaged for every metal size. Figure~\ref{fig:supp_structure} shows that the smallest clusters have little bulk-like FCC/HCP interior and display strongly size-dependent ordering, whereas the larger clusters form an increasingly ordered core and follow a smoother size sequence. Abrupt changes in these observables also coincide with the caloric transition intervals for most sizes, suggesting that changes in caloric parameters are associated with structural rearrangements and shell closure. 

We therefore selected $g=3043$ as the common lower boundary of the smooth asymptotic branch for both metals. Sizes below this threshold were retained in Figure~\ref{fig:supp_thermo} to show the departure from the model, but were excluded from parameter estimation and model-selection statistics. Consequently, values of $a_g$ and $L_g(T)$ below $g=3043$ in the growth calculations should be interpreted as a smooth-envelope extrapolation, not as a fit to magic-number or structure-specific oscillations of individual small clusters.

